\documentclass[preprint,12pt]{elsarticle}

\usepackage{amssymb}
\usepackage{amsmath}
\usepackage{svg}
 \usepackage{array,multirow,graphicx}
 \usepackage{float}
 \usepackage{tabularx}
 \usepackage{subfigure}
 \usepackage{caption}

\usepackage[hidelinks]{hyperref}
\biboptions{numbers,sort&compress}

\journal{Journal of Energy Storage}

\begin{document}

\begin{frontmatter}



\title{A continuous multiphase model for liquid metal batteries}


\author[inst1]{Omar E. Godinez-Brizuela*}
\ead{omar.e.g.brizuela@ntnu.no}
\affiliation[inst1]{organization={Norwegian University of Science and Technology},%Department and Organization
            %addressline={}, 
            city={Trondheim},
            postcode={7030}, 
            %state={},
            country={Norway}}

\author[inst2]{Carolina Duczek}
\author[inst2]{Norbert Weber}
\author[inst1]{Kristian E. Einarsrud}

\affiliation[inst2]{organization={Helmholtz Zentrum Dresden-Rosendorf},%Department and Organization
            addressline={Bautzner Landstr. 400}, 
            city={Dresden},
            postcode={01328}, 
            country={Germany}}

\begin{abstract}
Liquid metal batteries (LMBs) are a promising alternative for large-scale stationary energy storage for renewable applications. Using high-abundance electrode materials such as Sodium and Zinc is highly desirable due to their low cost and excellent cell potential. LMBs undergo multiple complex mass transport dynamics and as a result, their operation limits and other critical parameters are not fully understood yet. In this work, a multiphase numerical model was developed to resolve electrode and electrolyte components in 1D and simulate the discharge process of a Na-Zn battery including the interfacial displacement of the molten metal electrodes. The variation in electrolyte composition was predicted throughout the process, including the species distribution and its effect on the cell conductivity and capacity. Volume change and species redistribution were found to be important in predicting the maximum theoretical capacity of the cell when neglecting convective phenomena. 
\end{abstract}



\begin{keyword}
Liquid metal battery \sep Energy Storage \sep Multiphase flow \sep Na-Zn
\end{keyword}

\end{frontmatter}

\newpage
\section{Introduction}
 The transition to renewable energy sources comes with a number of unsolved technological challenges. In recent years, it has become clear that a significant reconfiguration of the technologies used for energy production and distribution will become necessary \cite{Zhang2021,weier_liquid_2017,hadjipaschalis2009overview}. Many renewable energy sources are intermittent in nature, meaning they are only available at certain times when certain conditions are met. This is combined with constant shifts in the energy demand and is only made more complex by unplanned disruptions in the resource supply chain.

Stationary, large-scale energy storage solutions have become necessary to address many of these challenges, but the existing technologies show significant limitations. Liquid metal batteries (LMBs) have emerged as a promising solution to many of the operational challenges in energy supply, offering high capacity and reliability \cite{Solheim2017}, with the ability to be produced from highly abundant, low-cost materials. This technology emerged as an alternative to other grid-level technologies available, such as pumped-hydro or large stacks of Li-ion batteries while addressing some of the disadvantages that make these systems infeasible, such as fire-related risks in the case of Li-ion, high cost \cite{Kim2013,li_liquid_2016}, and geographic infeasibility. 

LMBs consist of a pair of electrodes in liquid state separated by a molten salt electrolyte. These systems must thus operate at a high temperature, such that all of the components remain liquid throughout the operation cycle \cite{bradwell2010liquid,ding_room-temperature_2020,Kim2013}. The cell materials must be chosen such that the electrodes react electrochemically with the electrolyte, leading to the net charge or discharge of the cell.

\begin{figure}
    \centering
    \includegraphics[width=\textwidth]{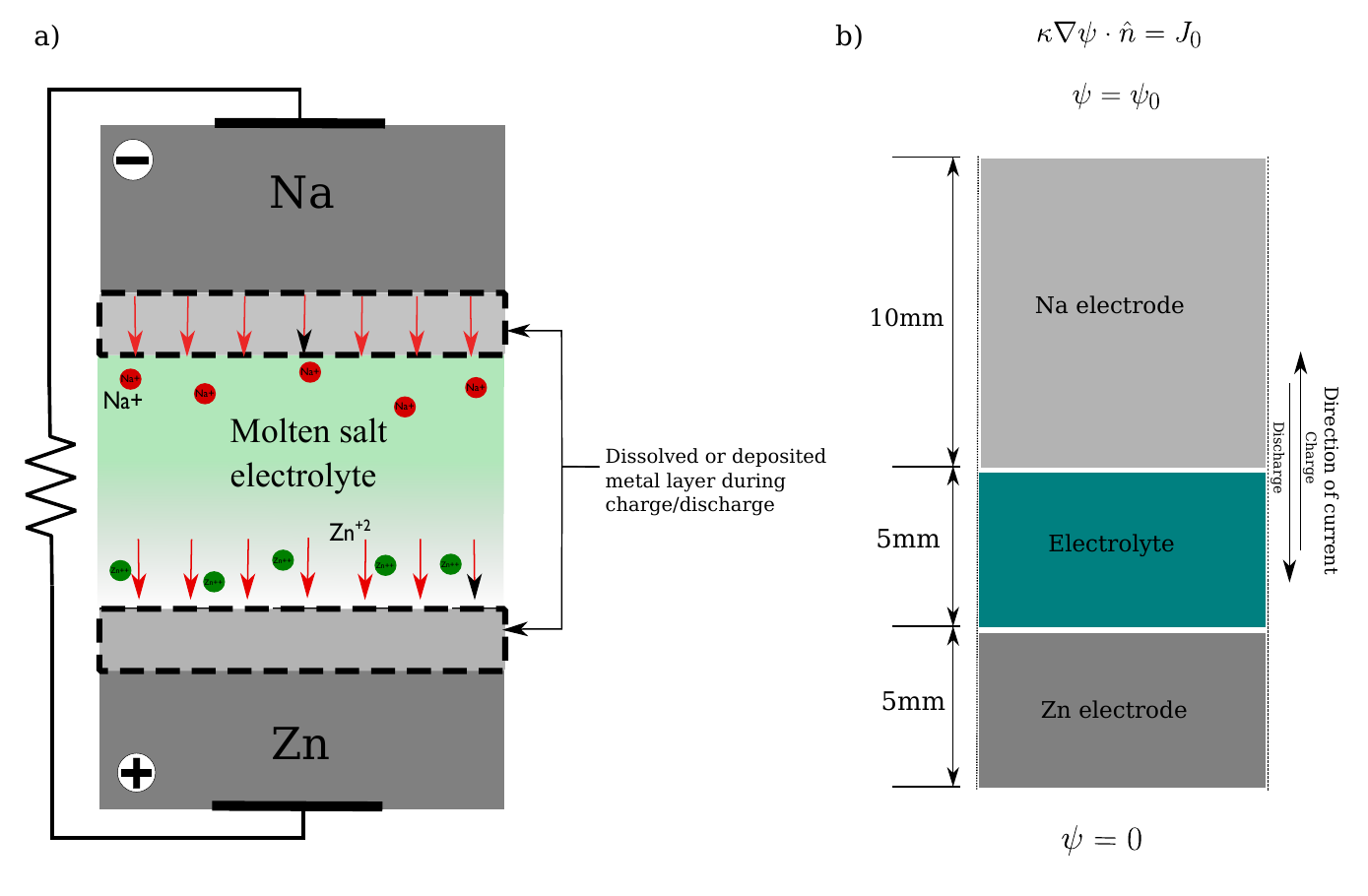} 
    \caption{Schematic figure of the liquid metal battery assembly, showing the moving boundary and the direction of the flux in red arrows during discharge (a). and Schematic showing dimensions of the domain and potential boundary conditions (b).}
    
    \label{fig:Diagram1}
\end{figure}

During the charge/discharge process, a metal displacement reaction occurs on the electrode interfaces, simultaneously dissolving one of the electrodes while depositing reduced metal on the other as shown in Figure \ref{fig:Diagram1} \cite{yin_faradaically_2018}. The components are assembled such that they are buoyant on top of each other due to their density difference \cite{herreman2020solutal,personnettaz2018thermally}. 

A ceramic membrane material can also separate the electrolyte to alleviate undesirable mass transfer between the electrodes \cite{yin_faradaically_2018, Xu2017a}, and to mitigate self-discharge.

Different selections of materials have been proposed in the past such as Li-Bi\cite{Newhouse2014,Ning2015}, Pb-Sb, Ca-Bi \cite{kim_calciumbismuth_2013} as examples. With the choice of electrode materials, often the aim is to use a pair of components with a favourable cell potential while maintaining a low cost. It must also be considered that the metals remain in a liquid state, so the electrode material choices are also constrained by the temperature at which the cell can operate \cite{Guo2021,ding_room-temperature_2020}. In general, the selection of materials for liquid metal batteries must target a high cell potential, low cost, and low melting point while also observing safety constraints. Most of the anode materials are suitable as they are relatively cheap and safe few exceptions such as Gallium-based batteries \cite{xie2020high} or those based on sodium-Potassium alloy \cite{ding2020room}. However, suitable materials for the anode are expensive to obtain and of low abundance \cite{entr2014report}.

Batteries based on the Na-Zn electrode pair have been proposed previously and appear to be a promising alternative \cite{Xu2016,Xu2017a,zhang_anode_2022}. Despite the high temperature needed to operate such a battery, the low cost and availability of Zn over other cathode elements make them attractive compared to other cell chemistries. This is combined with the reduced cost and availability of the molten salt components to be used as part of the electrolyte, and the wide availability of sodium \cite{Sun2019,Gong2020}.

Another important concern is the electrolyte materials, as there is a wide variety of molten salts that can potentially be utilized. In principle, the molten salt is typically chosen as a mixture of the corresponding chlorides of the electrode materials, with possibly other salts mixed in to alleviate problems of miscibility or corrosion. Addition for example of iodides, can significantly reduce the melting point, but comes with increased cost \cite{Kim2013}. The melting point of the electrolyte must be compatible with the operating temperature of the electrodes and should exhibit high ionic conductivity with low electronic conductivity\cite{liu_molten_2022,weier_liquid_2017,zhou_sodium_2022,Kim2013}.

Throughout the charge/discharge cycles, the composition of the electrolyte would be expected to change, and as such, the amount and initial composition of the salt would be linked to the overall capacity of the cell. Furthermore, the increase in thickness of the electrolyte layer can also dictate some of the operational limits of the cell, as this part of the assembly is the one with the highest resistivity, and accounts for most of the Ohmic losses \cite{weber_cell_2022,swinkels1971molten}. 

LMBs still require significant development to improve their efficiency, as they suffer from high self-discharge due to electrode miscibility, undesirable co-deposition, or corrosion \cite{Ning2015,zhou_sodium_2022,liu_molten_2022}% improve these.
. The particular self-discharge mechanism for Na-Zn systems has been described before \cite{zhang_anode_2022}.

In past years, several studies have aimed to understand other operational phenomena associated with LMBs, such as solutal convective effects \cite{herreman_solutal_2020,personnettaz_mass_2019,personnettaz2021effects}, and magneto-hydrodynamic effects, which also play a significant role in these cells, leading to electro-vortex flows and metal pad roll instabilities \cite{zikanov_metal_2015,herreman_numerical_2019,stefani_magnetohydrodynamic_2016}. It has been mentioned that these hydrodynamic effects are very important for the large-scale operation of the cell as interfacial instability may add another design constraint when aiming to build larger batteries \cite{Herreman2015,Kelley2018,horstmann_coupling_2018,benard_anode-metal_2021,Weber2013}. 
Understanding these convective phenomena is critical to avoid short circuits, and improve the mass transfer properties of LMBs.

Modelling solute transport in the electrolyte has been done in the past, focusing primarily on the estimation of concentration overpotentials in the cathode \cite{weber_numerical_2020,weber_cell_2022}. Further, ion transport in the electrolyte can also influence the cell performance to a large extent and attention and this has been quantified before for the case of Li-Bi cells \cite{duczek2023simulation} when fluid flow is neglected. One must also take into account that during the cycling of the cell, there is a simultaneous dissolution and electro-deposition process in the electrode surface. If there are significant density differences, the total volume of the layers will change, and this will lead to the electrode-electrolyte interfaces being displaced. This process is difficult to model accurately and thus this effect has not been fully explored in literature, and its impact on cell performance is not fully understood. Several previous works have modelled transport processes in other aspects of liquid metal batteries\cite{Newhouse2014,weber_cell_2022}, the detailed modelling of the volume change in the immiscible components is still not well understood. Previous models, investigating volume change in LMBs in contrast were  simplified or limited to analyzing static interfaces, at several layer thicknesses \cite{Newhouse2014,jiang2019effects}.

In this work, we aimed to create a simplified unidimensional model that can contain most of the mass transfer effects contained in the interaction between the molten salt electrolyte and the electrodes of a Na-Zn cell. Similar to the previous approach proposed \cite{duczek2023simulation}, but with further considering the variation in thickness of the different battery layers to find out to what extent the influences the operation and performance of the cell.
We approached this problem by developing a multiphase-flow method using a Volume-of-Fluid method for an arbitrary number of species, and capable of handling both miscible and immiscible sets of components \cite{Newhouse2014,weber_cell_2022}. For the purpose of observing the effects of purely of migration and diffusion in detail, more complex convection effects are neglected.

The main motivation of this model is to resolve the electrolyte volume change taking place as a result of the metal displacement reaction of the cell, while also resolving the species distribution within the electrolyte. With this information, we model the resulting changes in potential distribution and assess the change in ohmic losses through the operation of the cell. Additionally, we estimate how concentration gradients within the electrolyte can limit the available charge that can be drawn from the cell.

\section{Methodology}
\subsection{General model configuration}
The model was conceived as a three-layer assembly, representing the electrodes and the electrolyte layer as shown in Figure \ref{fig:Diagram1}. The dimensions chosen allow for the Na layer to be thicker, since it is of significantly lower density than the other materials, and thus an equivalent charge transfer would lead to a deeper erosion of the sodium layer than the Zn.

The system is modelled in these dimensions such that it is reasonable to neglect hydrodynamic effects and focus on the mass transfer processes that can be resolved unidimensionally.

The electrolyte layer is composed of a mixture of $\mbox{NaCl}$ and $\mbox{ZnCl}_2$ and is represented by a mixture of three species, representing the $\mbox{Na}^+$, $\mbox{Cl}^-$ and $\mbox{Zn}^{+2}$ ions. Initially, this mixture consists of mole fractions of 0.583, 0.25 and 0.167 for the Cl, Na and Zn ions respectively, to represent a mixture of 50\% $\mbox{NaCl}$ and 50\% $\mbox{ZnCl}_2$.

Constant current $J_0$ and constant potential $\psi_0$ conditions were applied to the domain from the electrode boundaries in the longitudinal direction, so as to observe the variation in concentration of the electrolyte components, as well as the formation or dissolution of electrodes at the  interface, displacing its location. The values for $J_0$ used were 100, 200, 500, 1000, 2000 and 5000 $\mbox{A/m}^2$ for the constant current cases, and voltage differences of 0.002, 0.004, 0.01, 0.02, 0.04, 0.1 V. These were repeated while inverting the sign of the current or voltage, to simulate the charge and discharge processes.

\subsection{Governing equations}
The proposed model uses different equations for miscible and immiscible components. Three immiscible components are defined first, representing the two electrodes and a region containing the electrolyte mixture, and their volume fraction ($\alpha_k$)is governed by equation \ref{vof} using the Volume-of-fluid method.
\begin{equation}
    \frac{\partial \alpha_k}{ \partial t}= R_k
    \label{vof}
\end{equation}
where k is an index representing the sodium electrode, the zinc electrode, or the electrolyte layer. $R_k$ is the reaction rate as a result of the current transfer at the interface. This term is calculated using Faraday's law as shown in equation \ref{Faraday}.

\begin{equation}
R_k=\nabla \cdot \sum_{i \neq k} \frac{ \alpha_k \mathbf{j} \delta_{k,i} \rho_k}{\eta_{k,i}N_A e_c M_k}
\label{Faraday}
\end{equation}
where $e_c$ is the elementary charge, $N_A$ is the Avogadro number, $\eta_{k,i}$ is the number of electrons transferred per mol at the electrode-electrolyte interface against phase $i$ and $\delta_{k,i}$ is an interface indicator function, defined as in Equation \ref{interface-marker}, where  $\alpha_i$ is the volume fraction of the phase $\alpha_k$ reacts with. $\mathbf{j}$ is the local current density at the electrode surface, defined as $\mathbf{j}=-\kappa \nabla \psi$ , where $\kappa$ is the local conductivity.  

\begin{equation}
\delta_{k,i}= \frac{\alpha_{k} \nabla \alpha_{i} - \alpha_{i} \nabla \alpha_{k}}{|\alpha_{k} \nabla \alpha_{i} - \alpha_{i} \nabla \alpha_{k}|}
\label{interface-marker}
\end{equation}
The molar concentration of each of the immiscible components is retrieved by dividing it by the corresponding molar volume:
\begin{equation}
c_k=\frac{\alpha_k \rho_k}{M_k}
    \label{molarConcentration}
\end{equation}

Within the electrolyte layer, The Nernst-Planck equation governs the transport of each of the ionic components. The molar concentration of each ion species is given by equation \ref{flux}.

\begin{equation}
    \frac{\partial c_i}{\partial t} = - \nabla \cdot \mathbf{N_{i}} + \sum_k\frac{\partial c_k}{\partial t}
    \label{flux}
\end{equation}

where the flux of i, $\mathbf{N_{i}}$ is given by the combination of diffusion, and migration, as shown in Equation \ref{nernst-planck}. The last term is the net reaction flux at the electrodes, where $c_k$ is the molar concentration of the k-th reacting electrode, thus the final term ensures the molar reaction rate appearing in the electrolyte matches the reaction rate of the electrode surface.

\begin{equation}
\mathbf{N_{i}}=-D_i (\nabla c_i )-\frac{D_i e_c z_i}{k_B T} c_i \nabla \psi  
\label{nernst-planck}
\end{equation}

where $D_i$ is the corresponding diffusion coefficient, $z_i$ is the ion valency, $k_B$ is the Boltzmann constant. $\psi$ is the electric potential, given by equation \ref{potential}, which ensures charge conservation in the conversion from ionic current to electronic current and vice-versa as the charge crosses the interface between the electrolyte and the electrode. 

Equation \ref{potential} is obtained by adding up the Nernst-Planck equation for each ionic species and converting each concentration into its equivalent charge density. Since the double layer is much thinner than the thickness of the electrolyte, the system is assumed to be electroneutral, thus cancelling out the accumulation of charge, and resulting in equation \ref{potential} \cite{newman2021electrochemical,duczek2023simulation}. 

\begin{equation}
\nabla \cdot  ((\kappa_e +\kappa_I  )\nabla \psi)= - \nabla \cdot \sum_i N_A e_c z_i D_i \nabla c_i 
\label{potential}
\end{equation}

The quantities $\kappa_e$, are the volume fraction-averaged electronic conductivity, and $\kappa_I$ is the ionic conductivity of the electrolyte. The right-hand side of the equation accounts for the net diffusion current \cite{duczek2023simulation}.  
The potential equation, defined in equation \ref{potential} uses the combination of the electronic conductivity $\kappa_e$ and the ionic conductivity $\kappa_I$, which is given by the approximation shown in equation \ref{NE}. 

\begin{equation}
\kappa_I= \sum_i \frac{D_i e_c^2 N_A z_i^2}{k_B T} c_i
\label{NE}
\end{equation}

The mass density of the molten salt is calculated as a mole-fraction-weighted average of its local compositions, density contribution from each ion is obtained from regression of the data found in literature \cite{janz1975molten}, where the individual ion contribution to density has been determined to be as shown in Table \ref{table_density}.
the electronic conductivity is calculated  as Equation \ref{averagekappa}.
\begin{equation}
    \kappa_e =\sum_i \kappa_i \alpha_i
    \label{averagekappa}
\end{equation}
Where in practice the electronic conductivity is only non-zero in the cells that are occupied by the electrode.
\begin{table}[hp]
\centering
\begin{tabular}{c c c c c c}
   Component  & $\rho_i$ $[\frac{kg}{m^3}]$ & $M_i$ $[\frac{kg}{kmol}]$ & $D_i$ $[\frac{m^2}{s}]$ & $z_i$ & $\kappa_e$ $[\frac{S}{m}]$ \\
    \hline
     Na$^{+}$& 1570.00 & 23  & $6.82\times10^{-9}$ & +1 & 0\\
     Zn$^{+2}$&  2580.00  & 65  & $1.97\times10^{-9}$ & +2 & 0\\
     Cl$^{-}$&  2470.00  & 35  & $ 1.19\times10^{-9}$& -1 & 0 \\
     Na & 817.12  & 23  & 0 & 0 & $3.471\times 10^6$\\
     Zn & 6436.67  & 65  & 0 & 0 & $2.739\times 10^6$ \\
\end{tabular}
\caption{Physical properties of system components. Diffusion coefficients are approximated based on their collective ionic conductivity \cite{janz1975molten,sobolev2011database}}.
\label{table_density}
\end{table}
With this approach, we aim to model the interface displacement process taking place when applying current through the given assembly of materials. 
Since we are primarily focused on observing the integrity of the interfaces and the evolution of the concentration profiles, we neglect all hydrodynamic effects including buoyancy for the purpose of this study. Additionally, the charge transfer overpotential is assumed to be negligible, such that discontinuous potential distributions at the interface are not  modelled \cite{Weber2019}. This is justified as it is known that the liquid interfaces allow for almost unhindered charge transfer between the electrode and the electrolyte. 

As a consequence of using equation \ref{potential} for modelling the potential, we assume that all electronic current approaching the interfaces causes a net ionic current of equal magnitude on the opposite side. 

It is assumed for the purposes of this work that the phase change only takes place between the ion and its matching metal. Namely, $\text{Na}^+$ can only be released at the $\text{Na}$ interface, and the $\text{Zn}^{+2}$ ion can only be reduced at the Zn surface. Thus, no co-deposition is allowed, and the values of the transfer coefficient $\eta_{ij}$ are 1 for the $\text{Na}$-$\text{Na}^+$ reaction, 2 for the $\text{Zn}$-$\text{Zn}^{+2}$ reaction. The reaction term is not added when the components in question are not reactive, which in this case corresponds to the $\text{Cl}^-$ ion.

\subsection{Numerical model}
These equations were modelled using the formulation  of the Finite Volume Method based on OpenFOAM. The model is based on the OpenFOAM solver \textit{multiphaseInterFoam}, \cite{greenshieldsweller2022} which was modified to accept the solution of miscible species within the electrolyte layer, as well as interfacial phase exchange, following the magnitude of the current passing through the system in accordance to Faraday's law. 

The computational domain is discretized unidimensionally and in only the direction normal to the electrode surfaces. We discretized the domain in 1000 cells of equal size. Time derivatives are discretized using the first-order Euler scheme, the divergence operators in equation \ref{nernst-planck} are discretized using the second-order upwind scheme, while the Laplacian term in equation \ref{potential} was discretized using second-order central differences. 

These equations controlling the evolution of the immiscible layers are solved explicitly, while the species distribution within the electrolyte phase was solved implicitly.

\section{Results and discussion} 

\begin{figure}
    \centering
    \begin{tabularx}{\textwidth}{ c X }
    t= 0 s &
    \includegraphics[width=7cm]{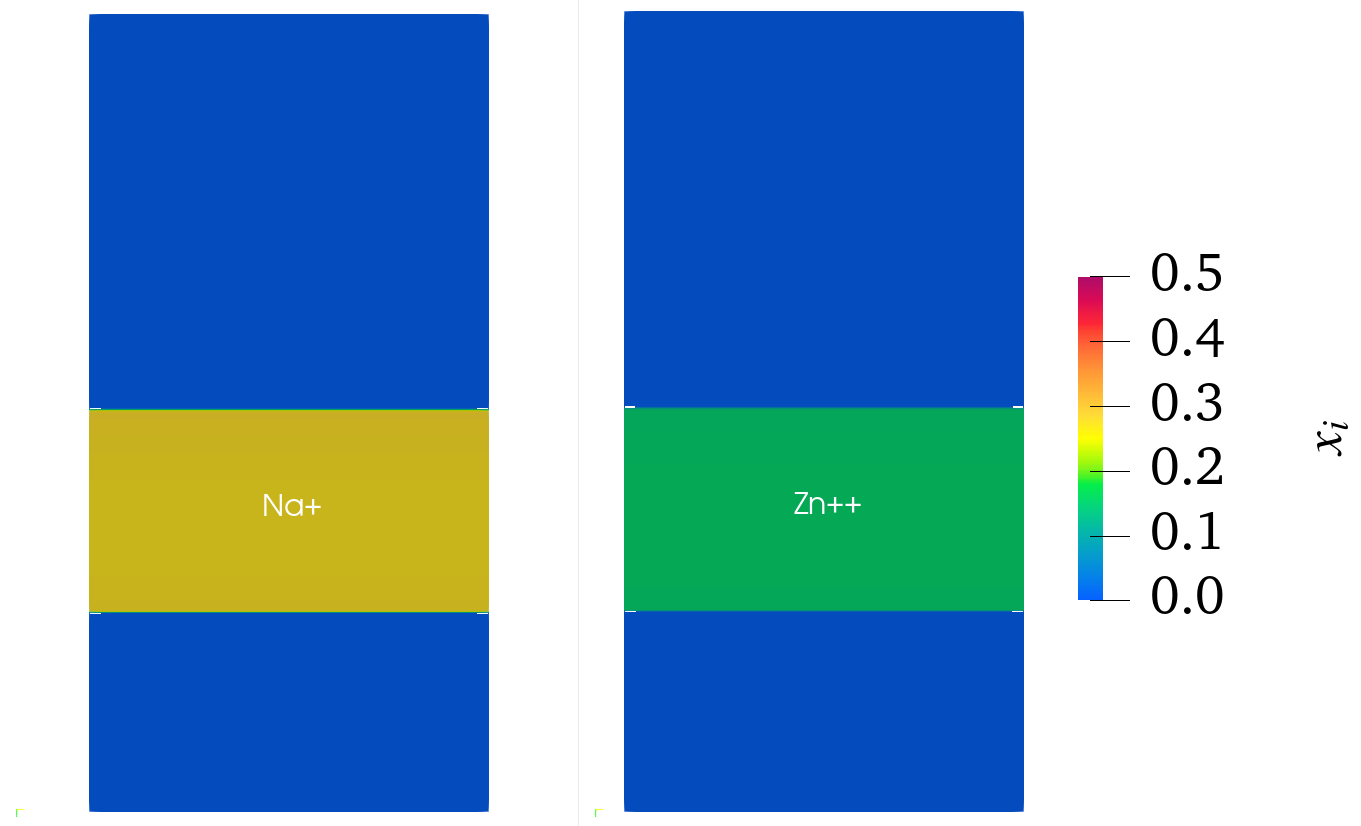} \\
    t= 20,000 s & \includegraphics[width=7cm]{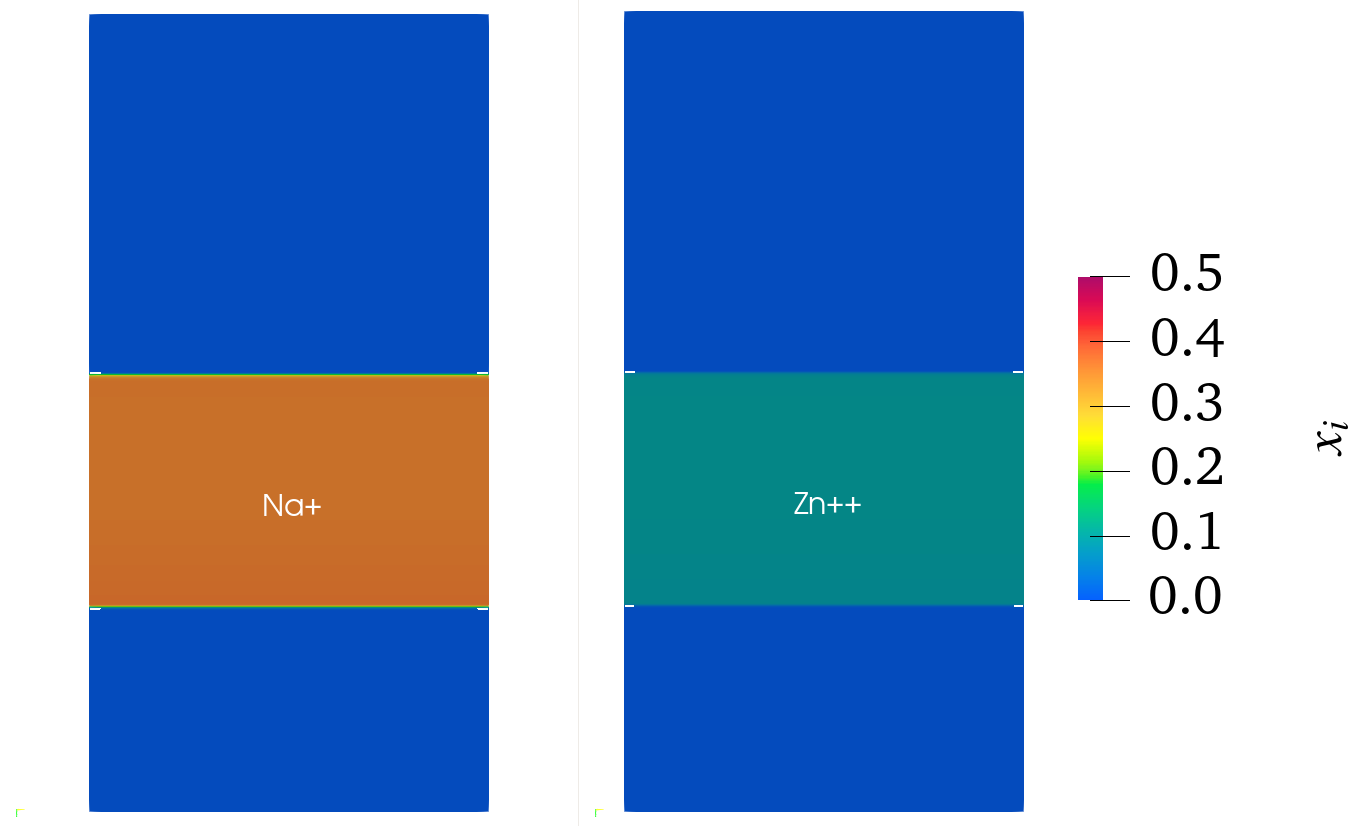} \\
     t= 40,000 s & \includegraphics[width=7cm]{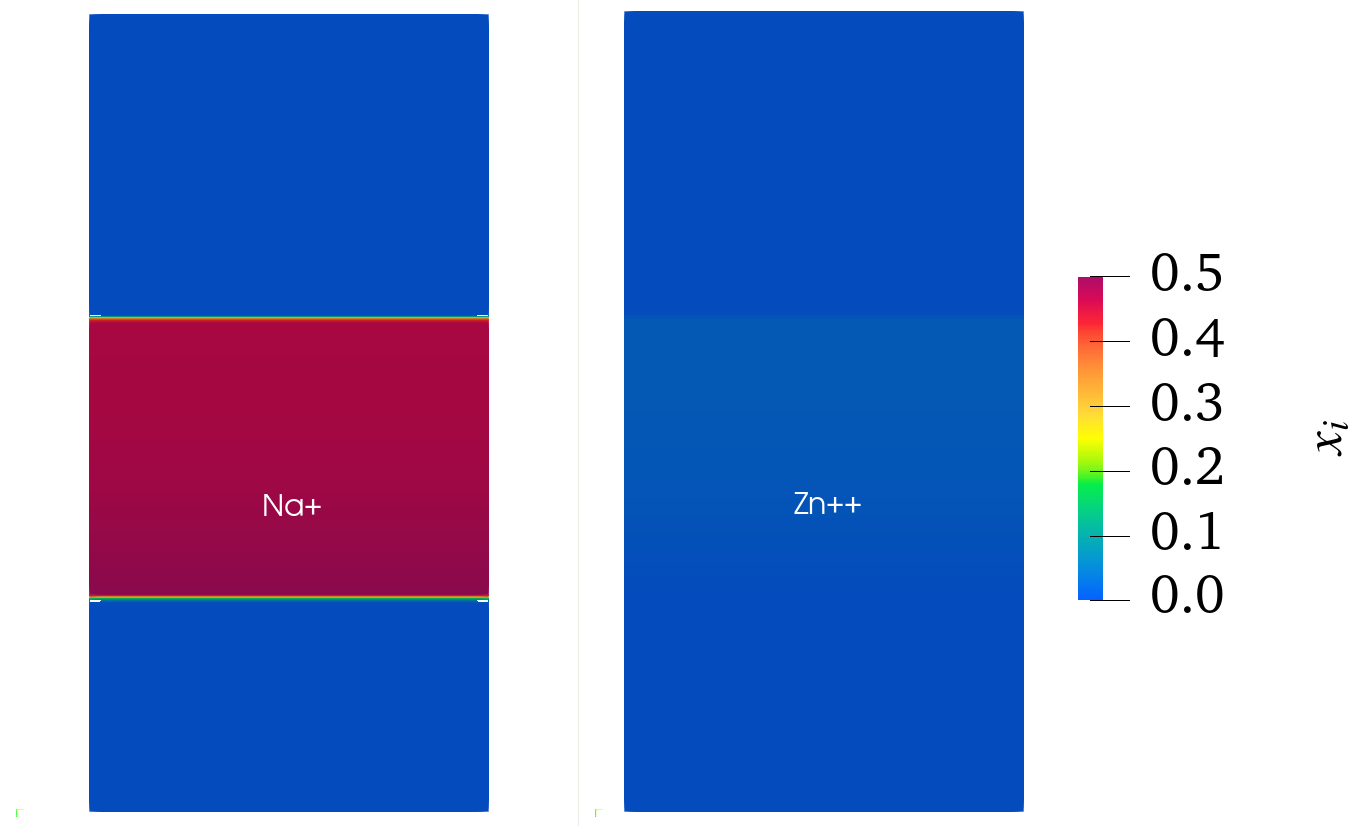}
    \end{tabularx}
    \caption{2D visualization of sodium and zinc ion concentration variation during the discharge process at $100$ $A/m^2$.}
    \label{fig:2Dfigures}
\end{figure}

As an initial example of the discharge process, it is possible to visualize the enrichment of the electrolyte in sodium ions while the existing zinc is being deposited in the bottom layer in Figure \ref{fig:2Dfigures}, showing the evolution of the composition until it reaches its fully discharged state.  
It can be noticed that the total electrolyte layer thickness increases during discharge due to the difference in density between the initial condition and the discharged state. This is also possible because the molar volume of sodium is much lower than that of zinc, and thus for a given current the sodium interface advance is significantly higher as shown in equation \ref{Faraday}.   

\begin{figure}
    \centering
    Initial state \\
    \includegraphics[width=8cm]{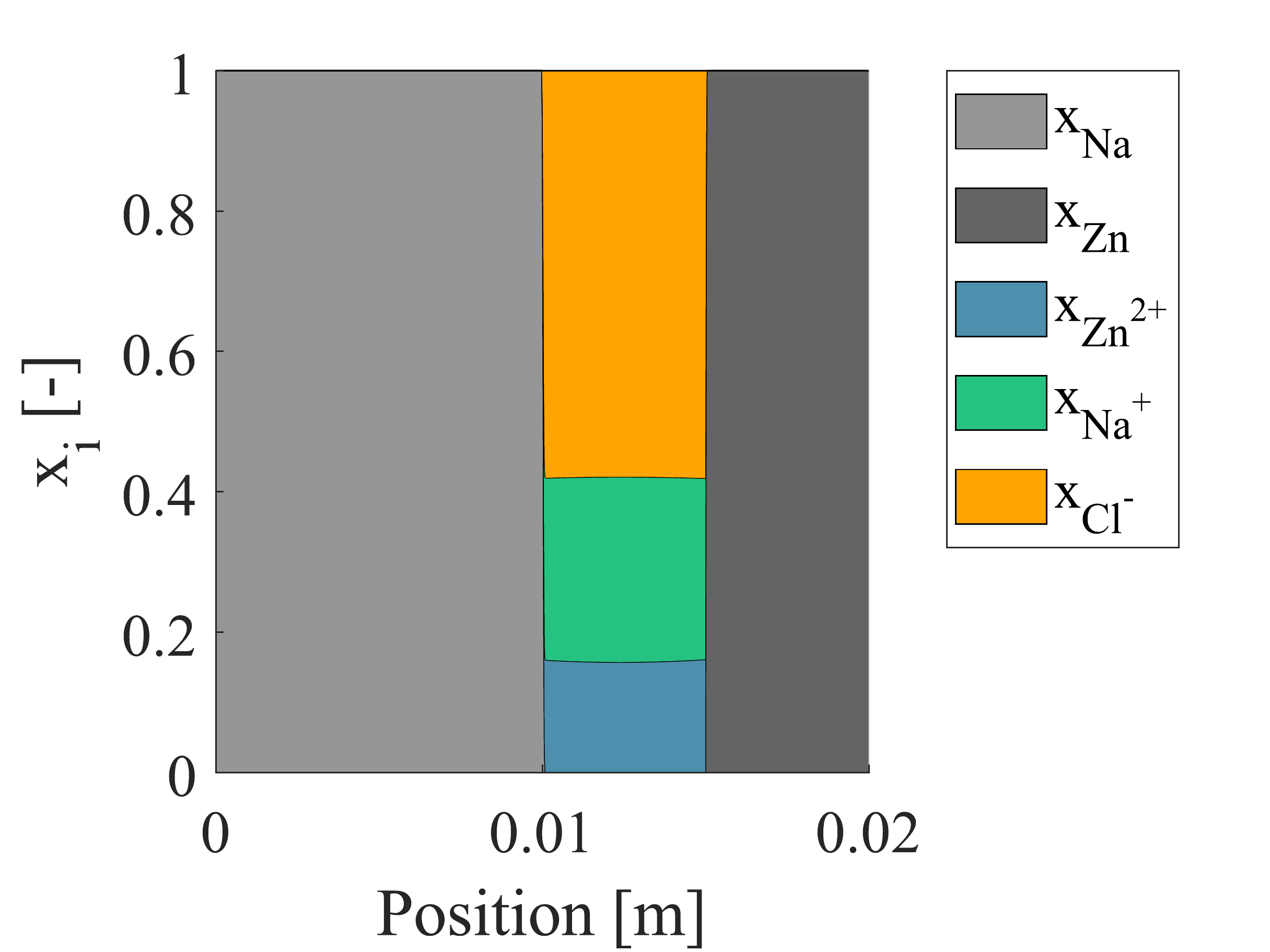}
    \\
    \vspace{1cm}
    \begin{tabularx}{\textwidth}{ c | X X }
    \hspace{1cm} & \hspace{1cm}After charge  & \hspace{1cm}After discharge\\
    \hline\\
     100 $A/m^2$ &
         \includegraphics[width=5cm]{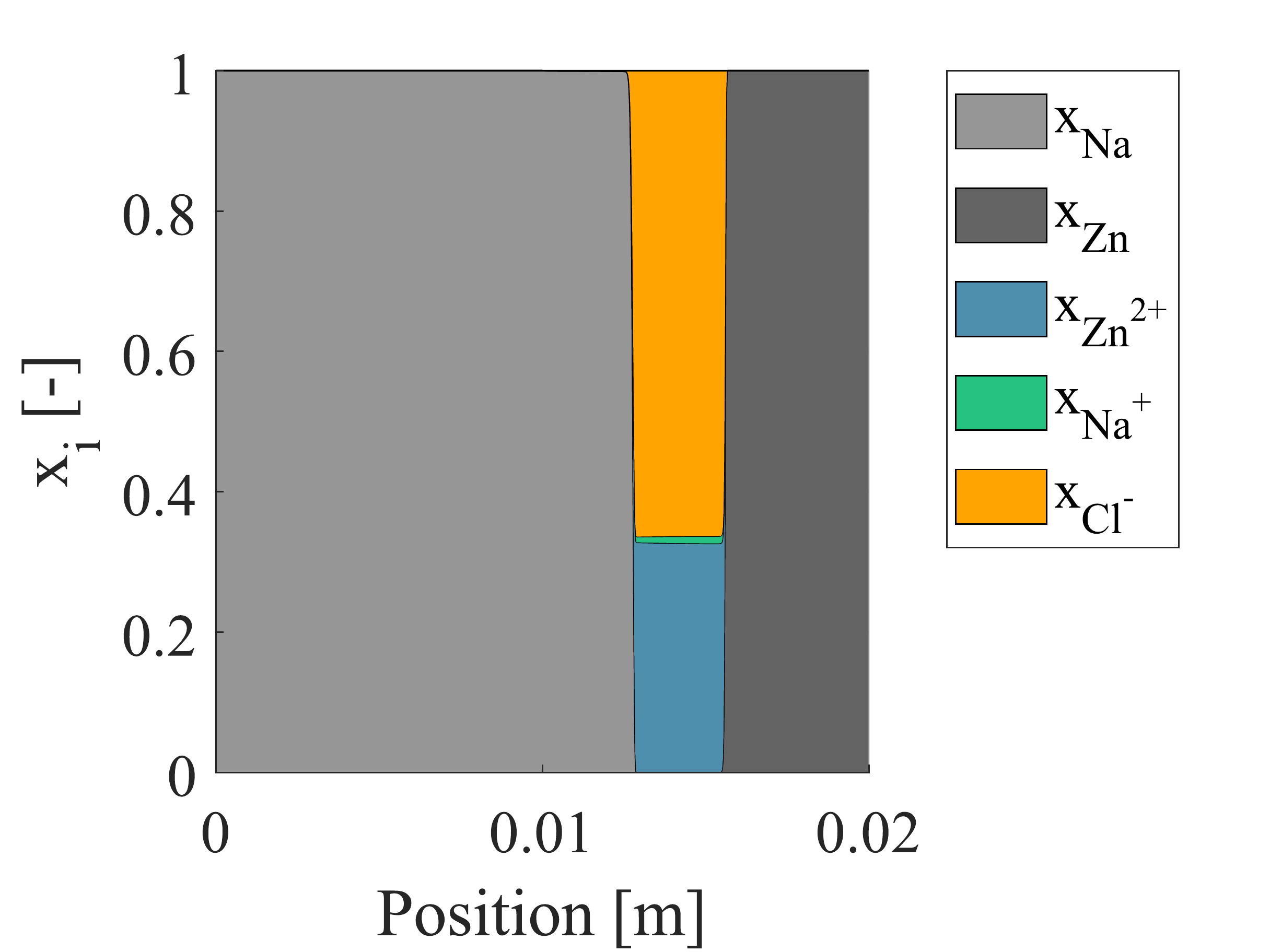}  & \includegraphics[width=5cm]{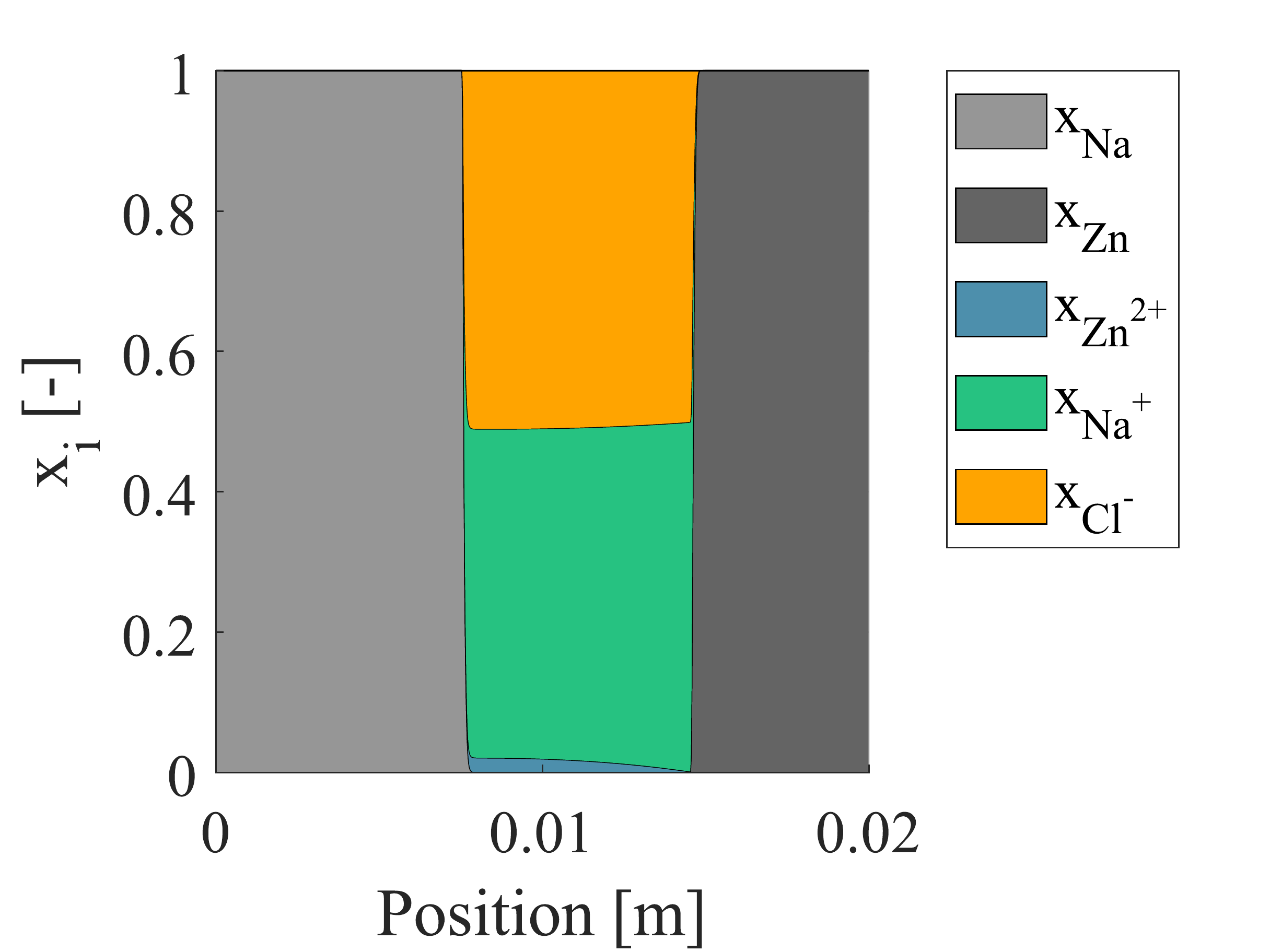} \\
     2000 $A/m^2$ & 
        \includegraphics[width=5cm]{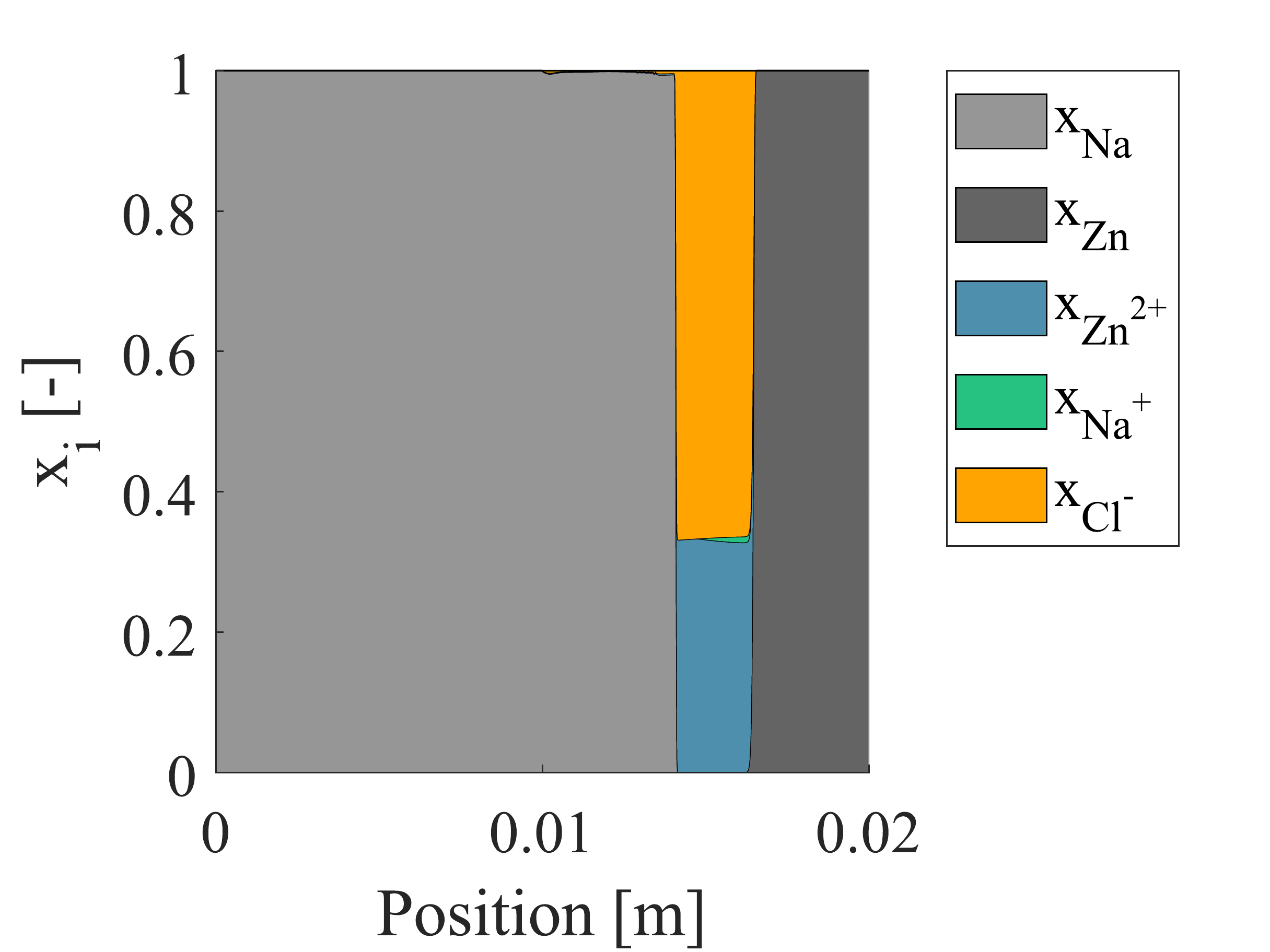} & \includegraphics[width=5cm]{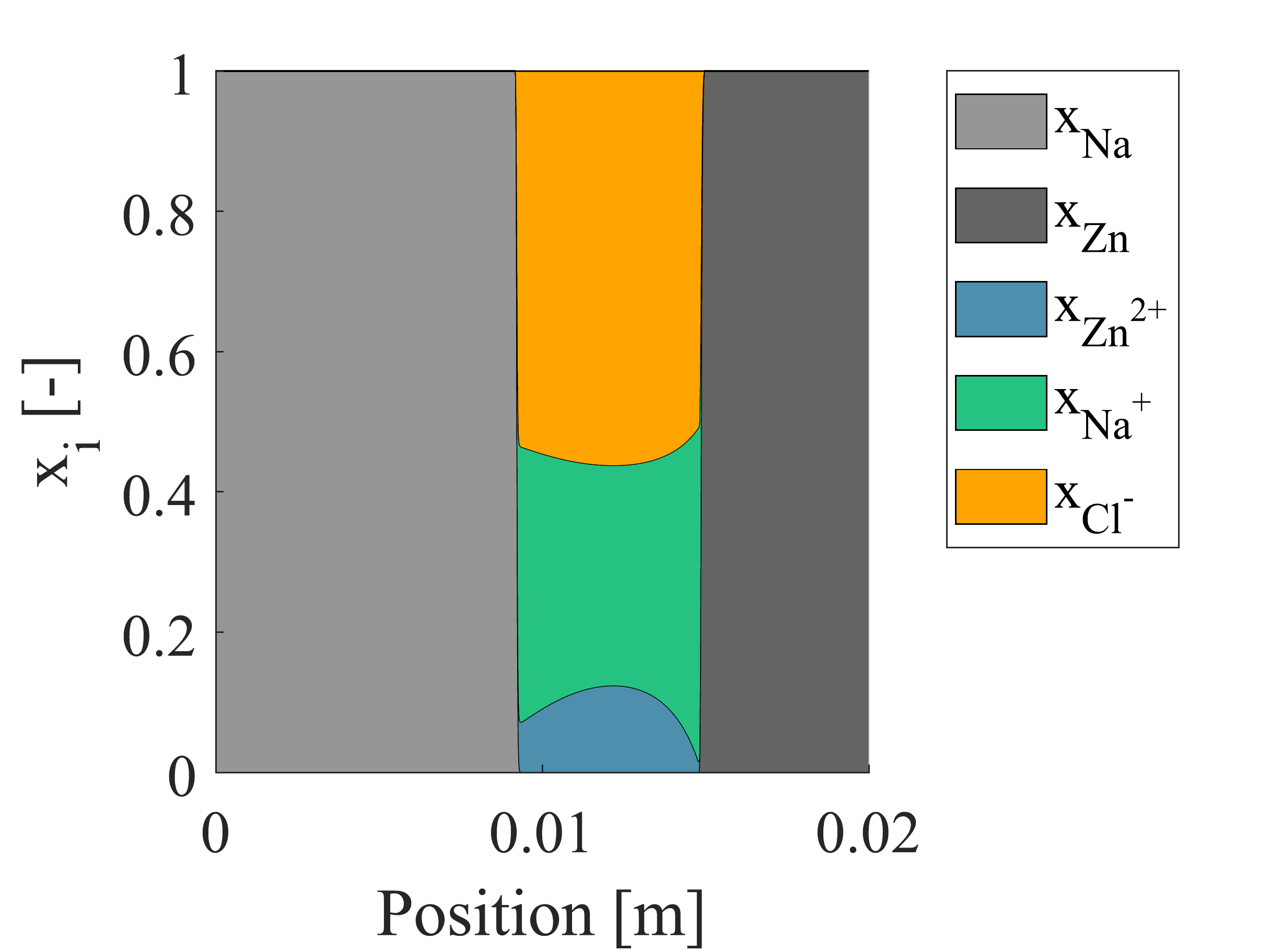}
    \end{tabularx}
    \caption{Area plot of the battery components (presented horizontally) in the initial state of the simulation. The Left side contains the sodium electrode and the right side the zinc electrode. The stacked areas represent the mole fraction of each ion in the electrolyte (top). Final states at two different current densities are shown below after charge and discharge.} 
    \label{fig:Composition Maps}
\end{figure}

The resulting species distributions after charge and discharge at high and low current conditions are shown in Figure \ref{fig:Composition Maps} as stacked area plots. Here one can visualize the proportion of each component occupying space noticing the sodium and zinc electrode regions on the side and the overlapped areas showing the electrolyte components.

The initial state of the simulations is shown in the top half of Figure \ref{fig:Composition Maps}, showing the mole fraction at the initial state. After submitting this composition to charge or discharge the distributions obtained are shown in the bottom half of Figure \ref{fig:Composition Maps}. The simulation ended when the depositing electrode lost contact with its corresponding reacting ion. So during discharge, the simulations are ended when the $\text{Zn}^{+2}$ concentration drops to zero near the zinc interface.

It can be observed in the final distributions that the final states at high and low currents are different, as the species distribution in the electrolyte layer is linked with the magnitude of the migration flux, so at higher currents, stronger concentration gradients are observed, limiting the amount of charge that can be stored or extracted from the cell.

In the high current example shown in Figure \ref{fig:Composition Maps}, one can observe a non-monotonic mole fraction distribution in the remaining $\text{Zn}^{+2}$ after discharge. This is because of the combined effect of the reaction rate near the depositing interface, and the displacement of $\text{Zn}^{+2}$ created from the release of $\text{Na}^{+}$ at the Na surface. At lower current densities, the magnitude of these concentration gradients is much lower, ensuring a deeper depth of discharge as it takes longer for the electrode to deplete the available reacting ion.

\begin{figure}
\footnotesize
    \begin{subfigure}[Whole system]{
        \begin{tabularx}{0.8\textwidth}{r X X}
        \hspace{1cm} & \hspace{1cm} Constant current & \hspace{1cm} Constant potential \\
    Charge & \includegraphics[width=5cm]{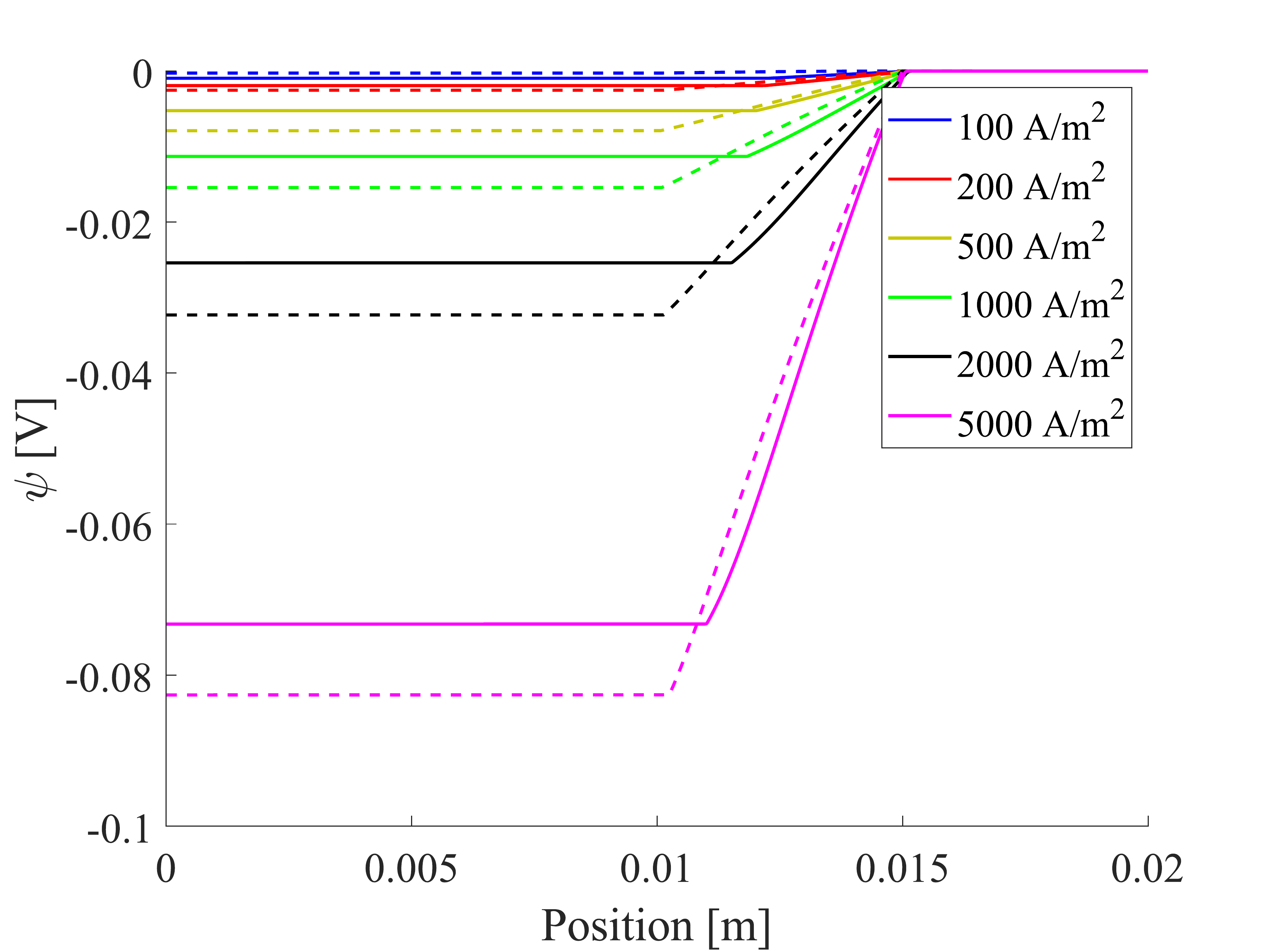} & \includegraphics[width=5cm]{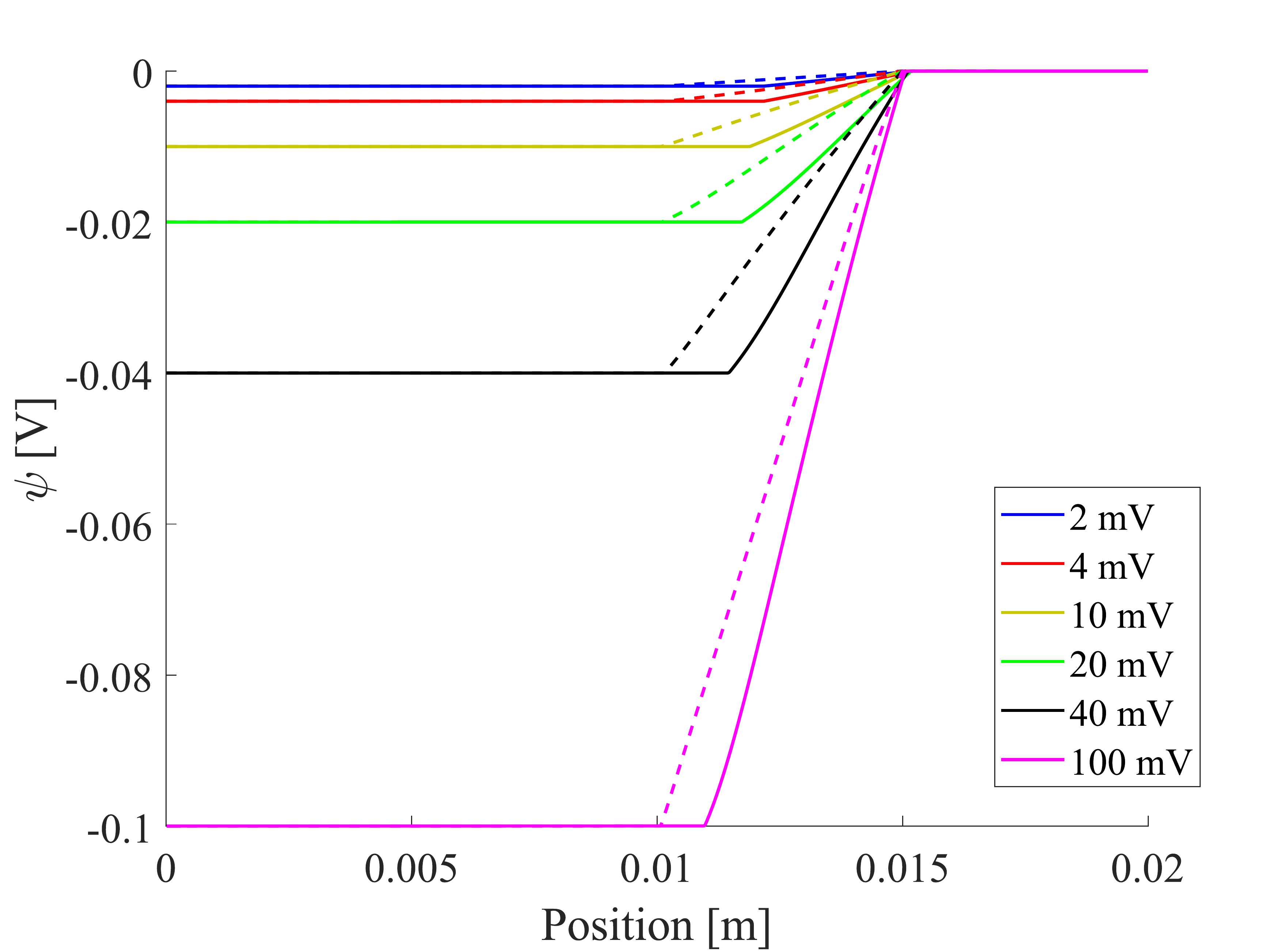} \\
        Discharge & \includegraphics[width=5cm]{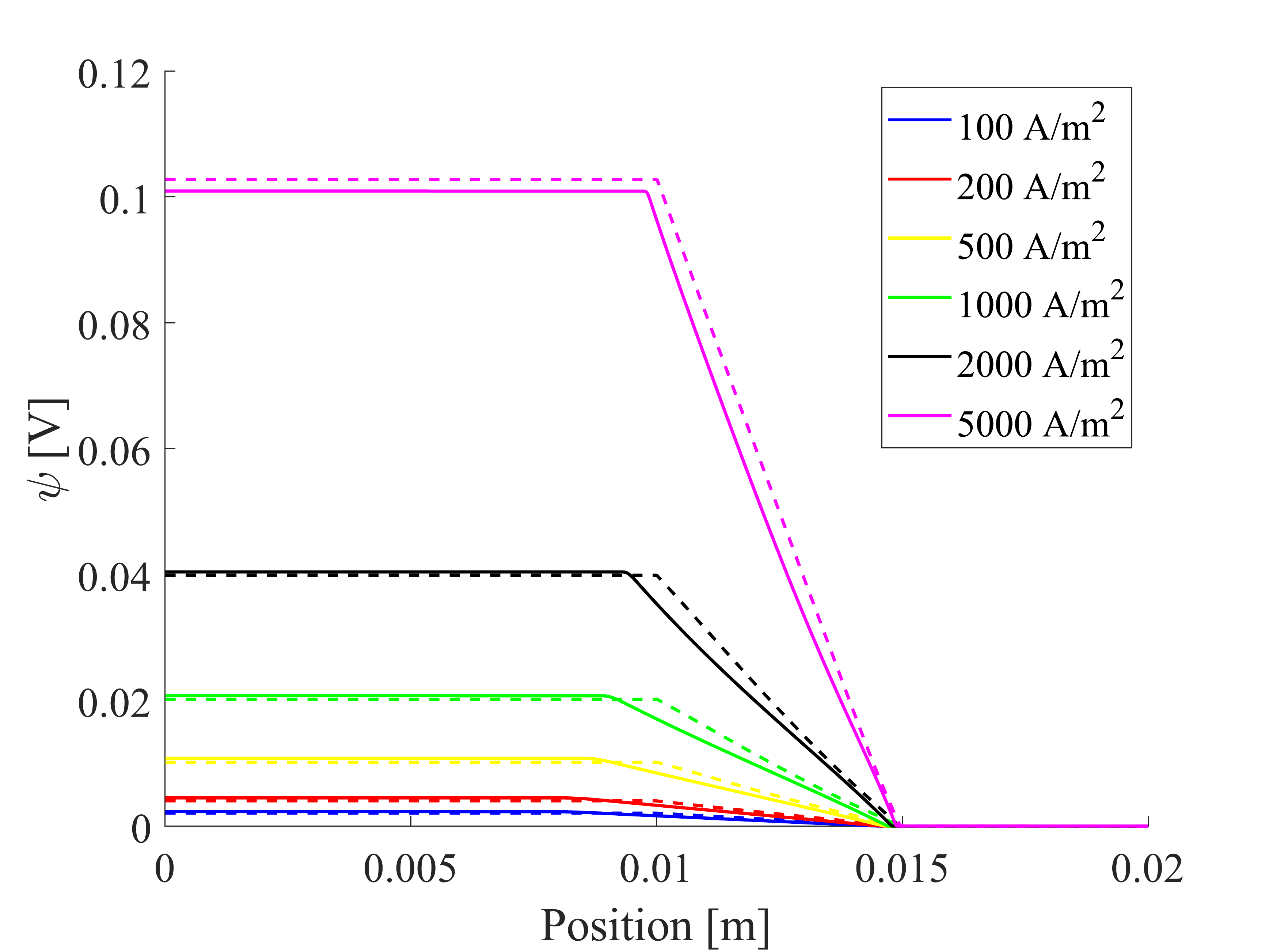} & \includegraphics[width=5cm]{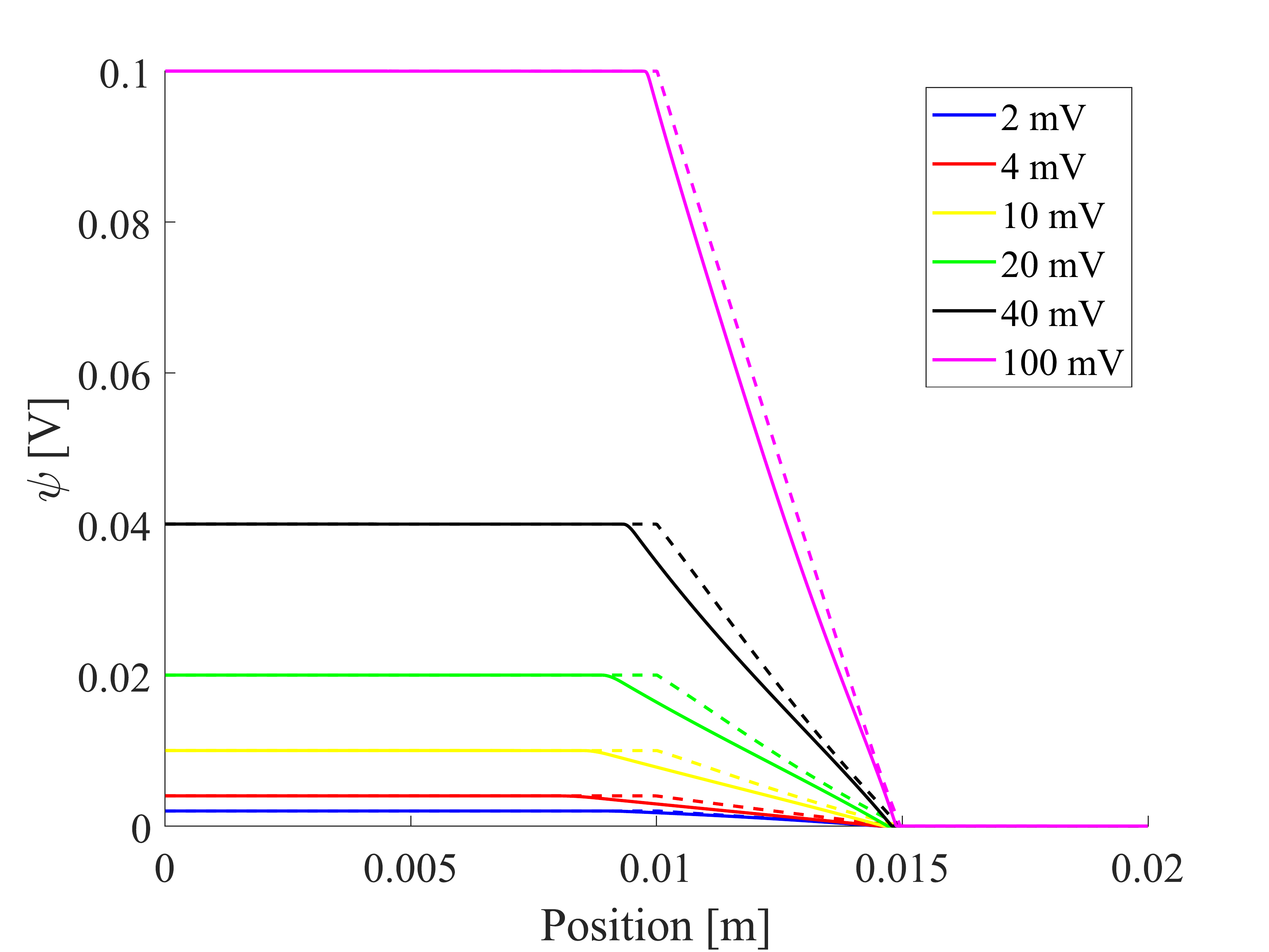}
    \end{tabularx}
    \label{fig:potential} }
\end{subfigure}\\

\begin{subfigure}[Within electrolyte]{\\
    \begin{tabularx}{0.8\textwidth}{r X X}\\
    \hspace{1cm} & \hspace{1cm} Constant current & \hspace{1cm} Constant potential \\
    Charge & \includegraphics[width=5cm]{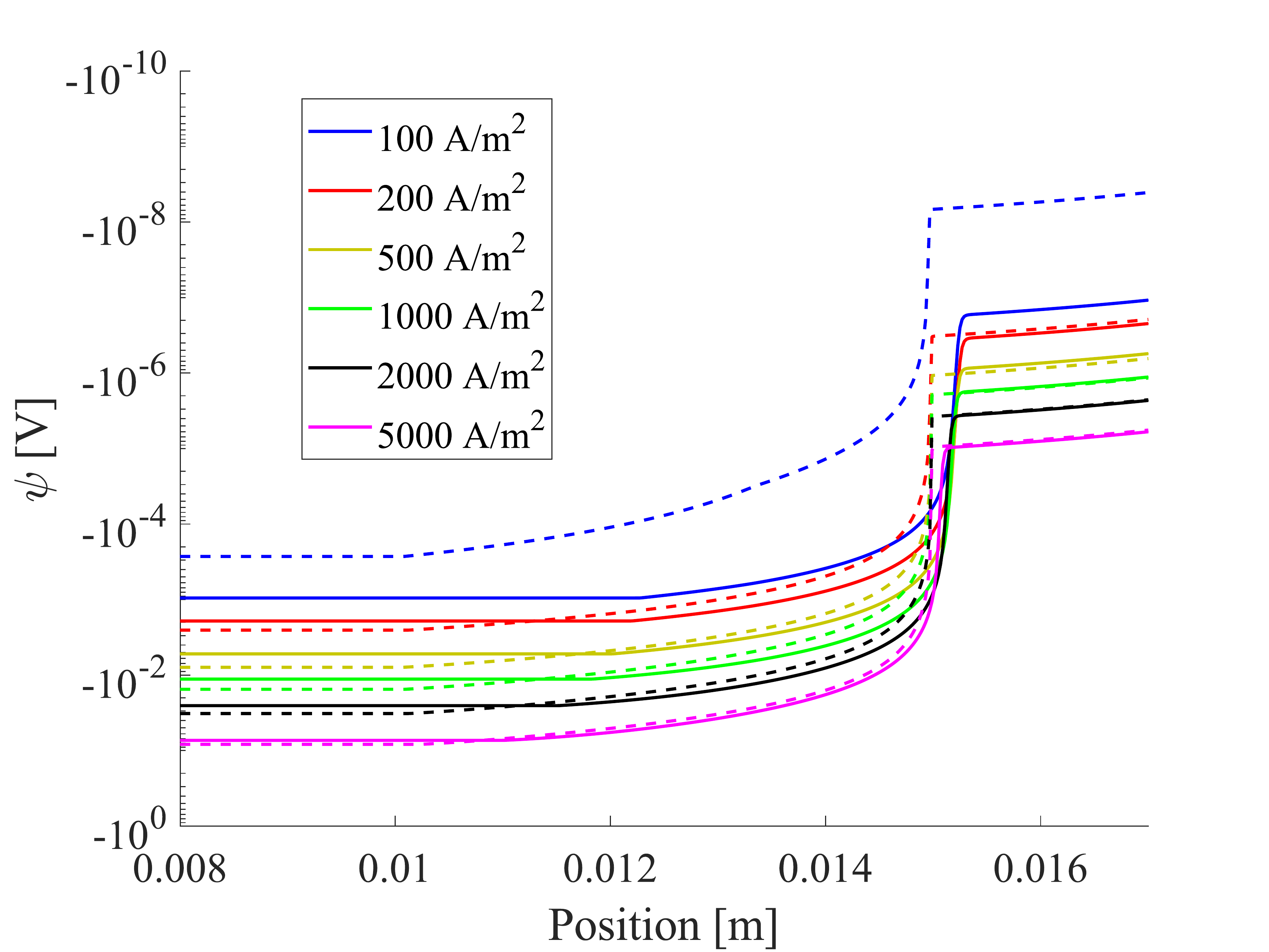} & \includegraphics[width=5cm]{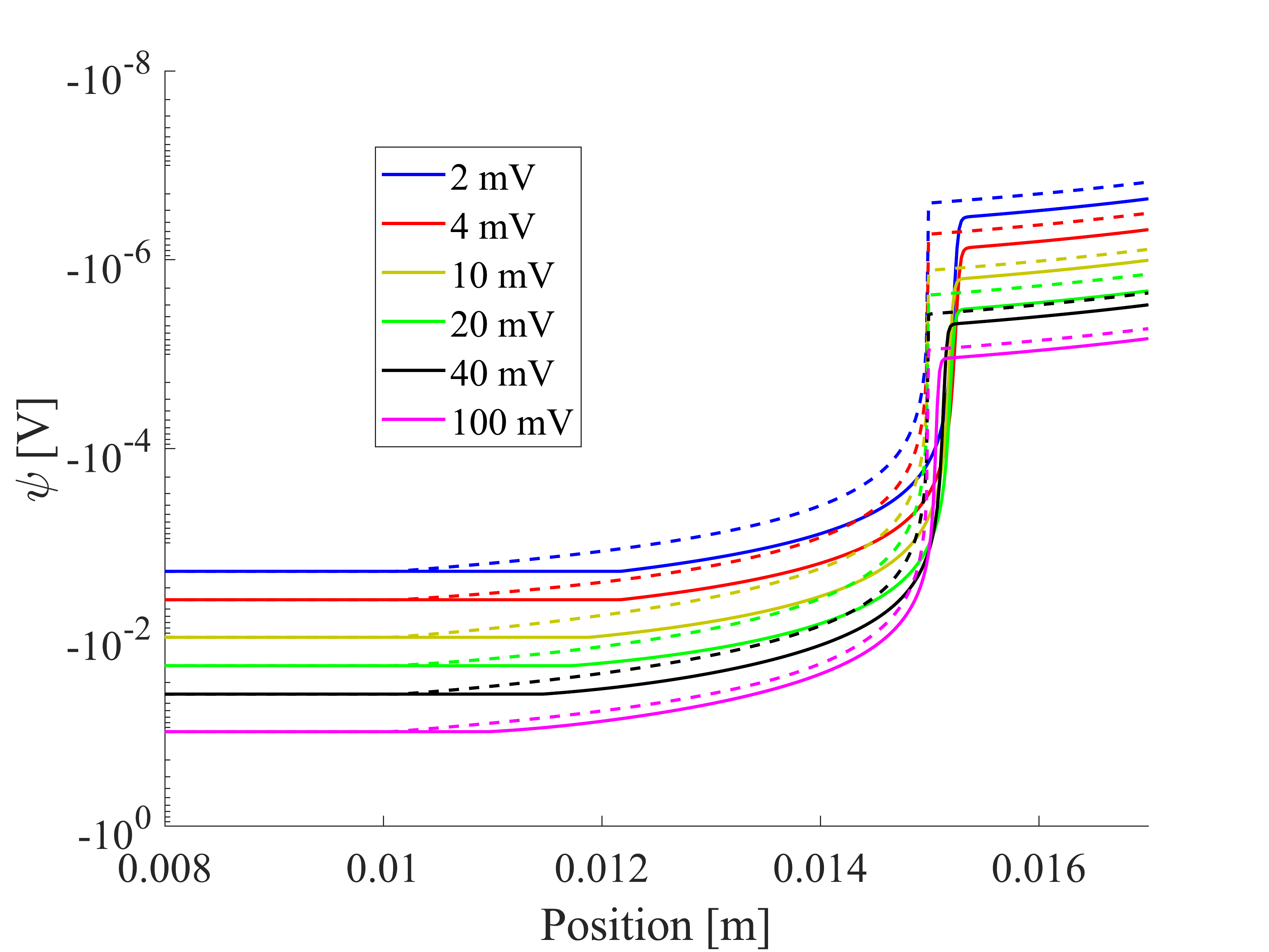} \\
        Discharge & \includegraphics[width=5cm]{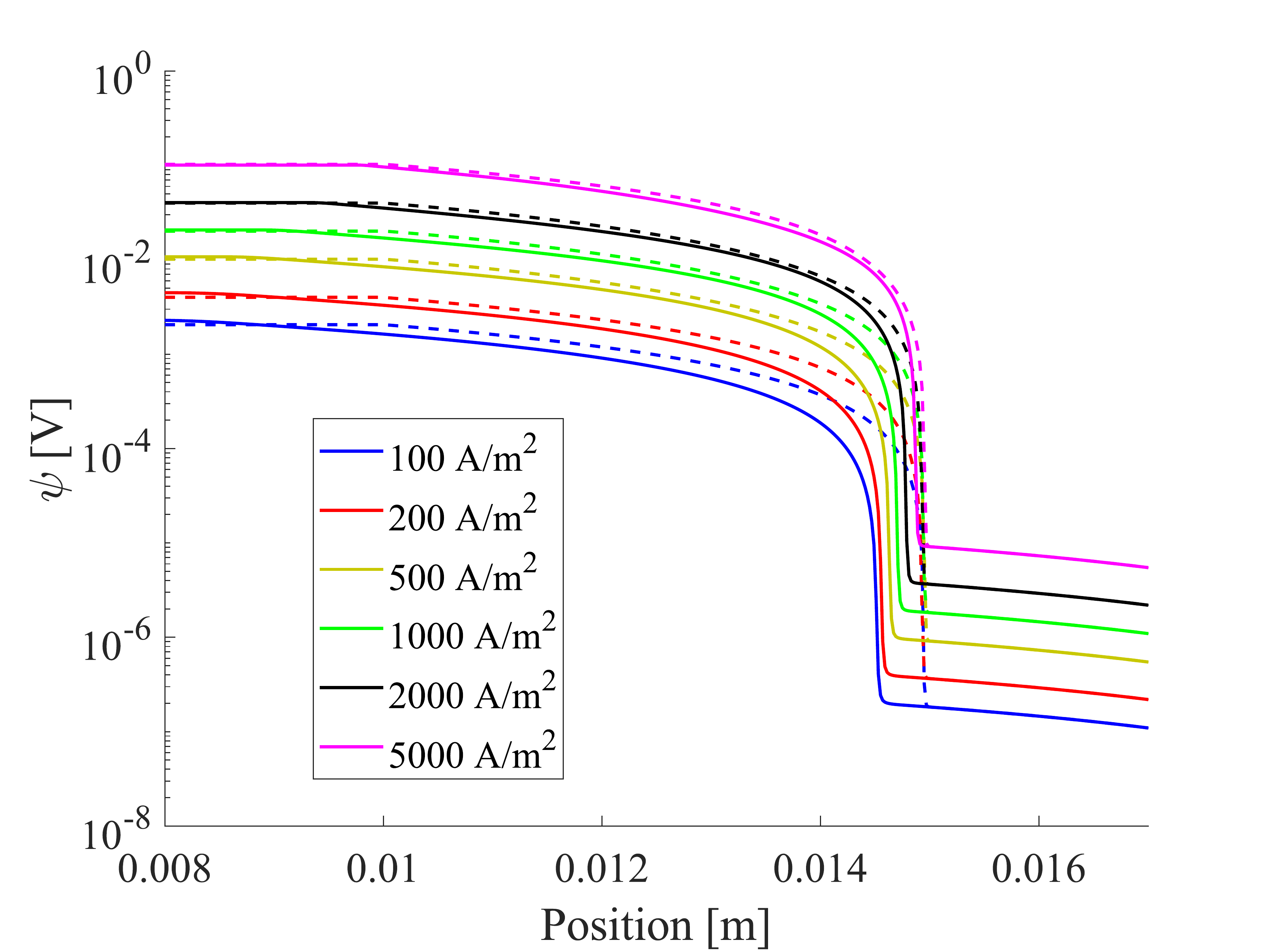} & \includegraphics[width=5cm]{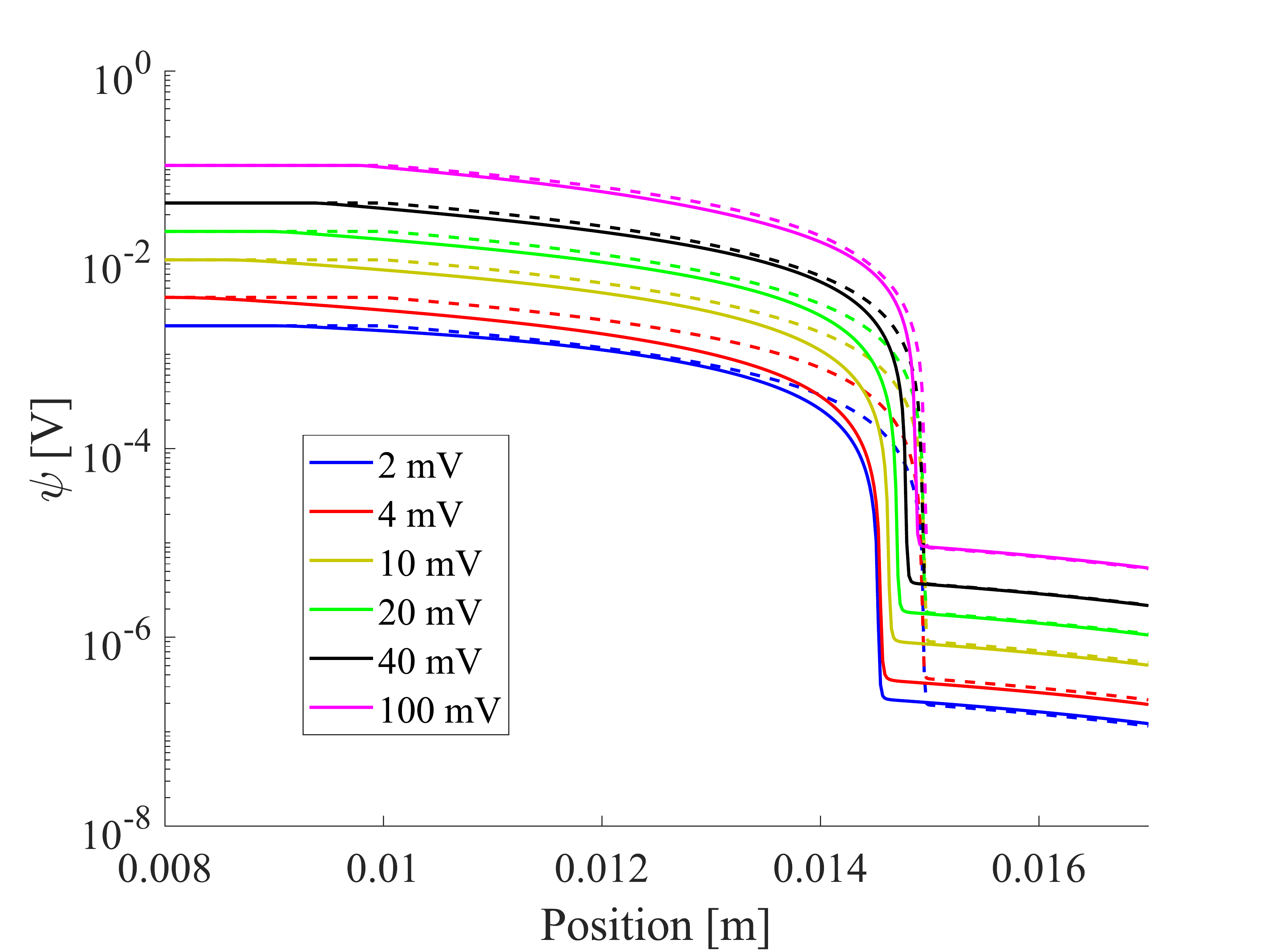}
    \end{tabularx}
    \label{fig:potential2}
    }
    \end{subfigure}

    \caption{Potential distributions across the cells (oriented horizontally). The dashed line is the initial state and the solid line is the final configuration.  The distribution across the whole system is shown in part (a) while the corresponding distribution within the electrolyte is shown zoomed in (b)}

\end{figure}

The potential distribution at initial and final conditions for each case is shown in Figure \ref{fig:potential}. In all cases it is shown that the variation of the thickness in the electrolyte layer creates a modification of the potential field, resulting in increased total resistance as the layer becomes thicker, and the opposite when it shrinks. In the constant current cases the potentials are adjusted while maintaining an equivalent current through the cell, however, because the conductivity of the cell is also changing, the slope in the potential field also changes. 

The constant potential cases show that the difference in current is mostly perceived in the electrolyte layer, as the electrodes have significantly higher conductivity, and thus the variation in current is directly linked to the expansion of the electrolyte layer. The lower current and applied voltage cases show a larger change between the initial and final configurations due to the more stable charge/discharge process.

A closer look at the potential distribution in the electrolyte can be seen in Figure \ref{fig:potential2}, where the shift in the distribution due to the change in thickness can be clearly seen. Additionally, it is possible to see that the deeper discharge possible in the low current/low voltage cases results in greater variations in this shift between initial and final states. 

\begin{figure}
    \centering
        \begin{tabularx}{\textwidth}{c X X}
        \hspace{1cm} & Constant current & Constant potential \\
        Charge  & \includegraphics[width=5cm]{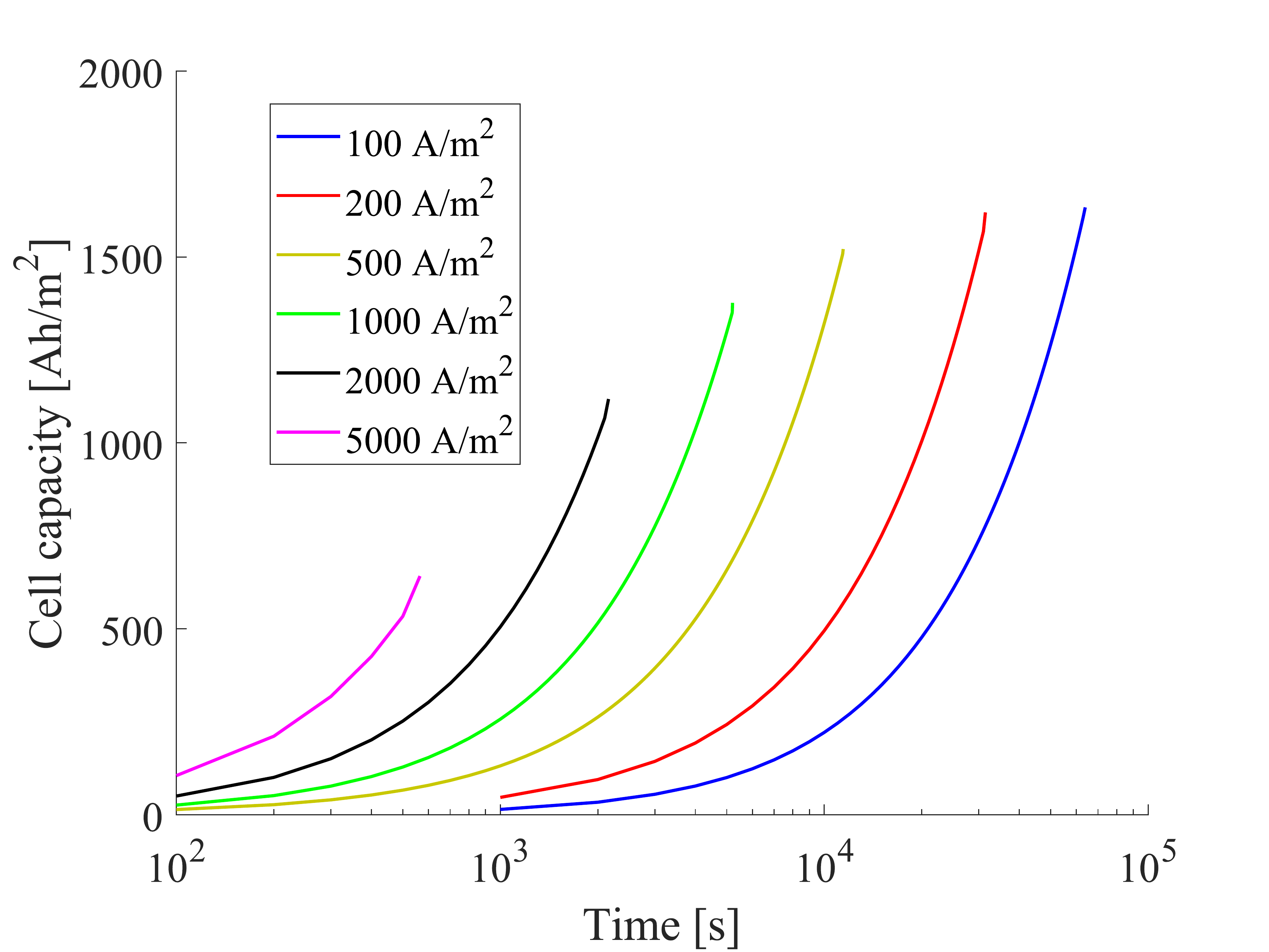} & \includegraphics[width=5cm]{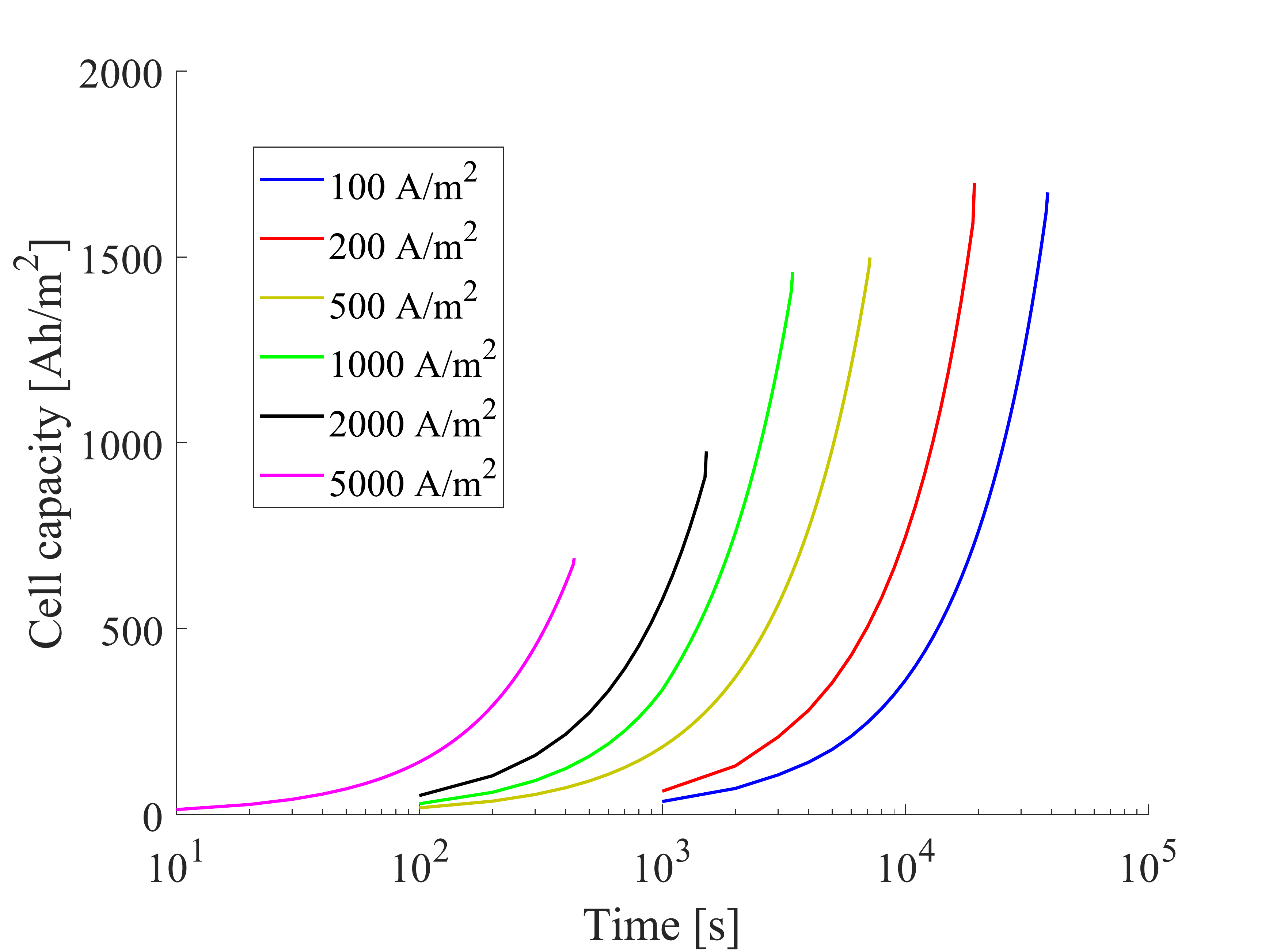} \\
        Discharge & \includegraphics[width=5cm]{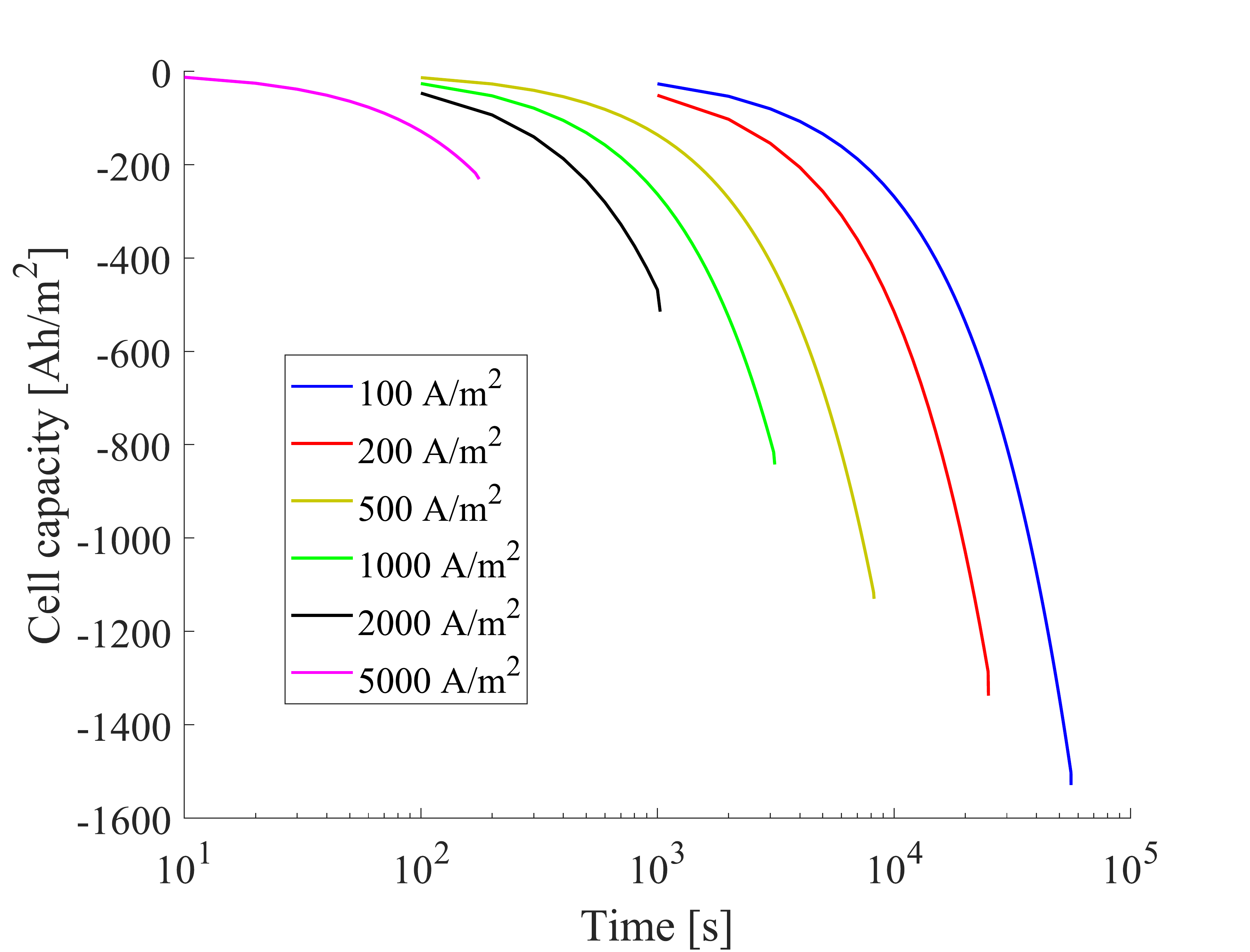} & \includegraphics[width=5cm]{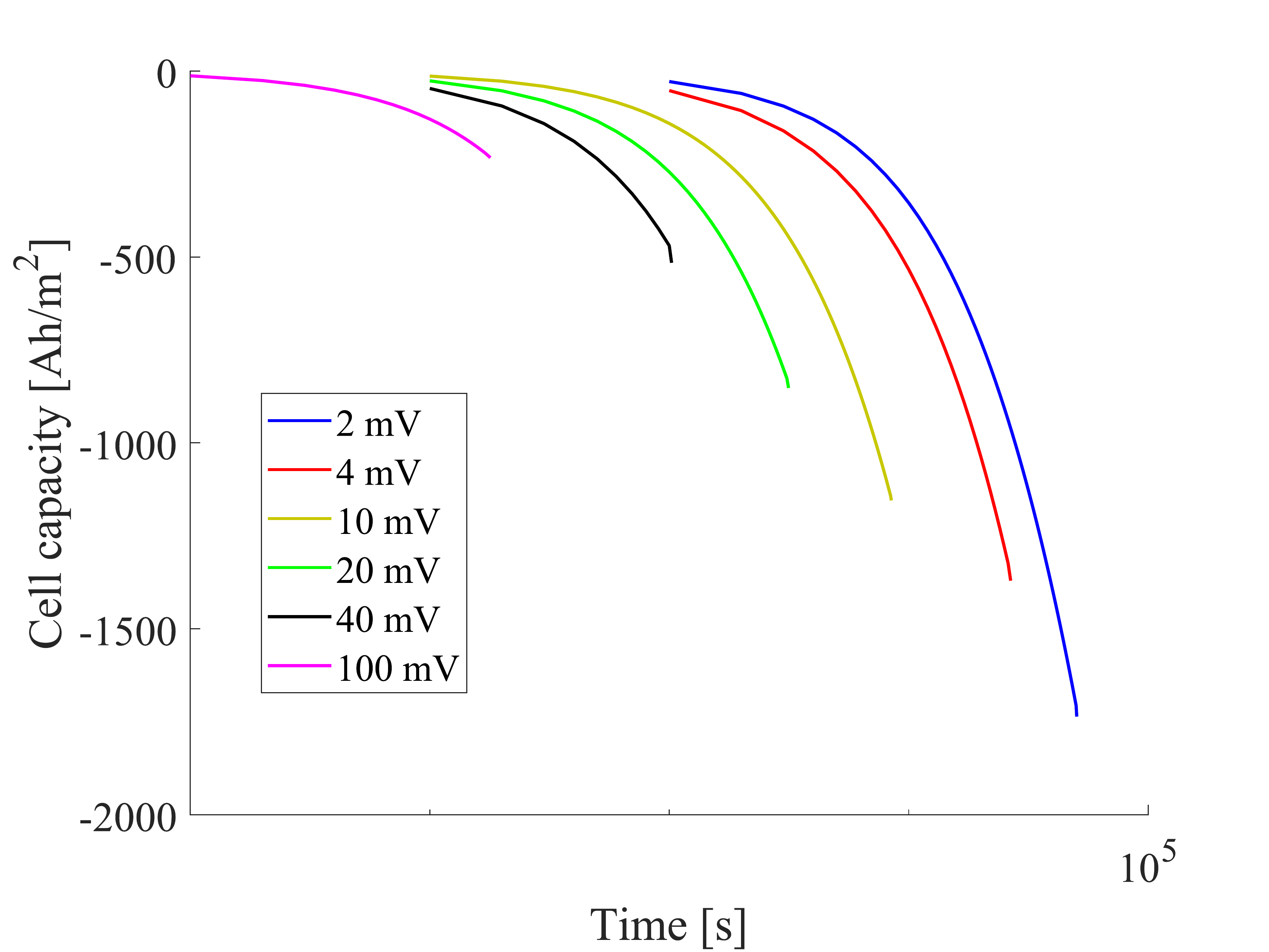}
    \end{tabularx}
    \caption{Cell capacity over time for each case.}
    \label{fig: Capacity vs time}
\end{figure}

The charge capacity stored or released by the system was calculated by integrating the current density through the anode boundary in time as shown in equation \ref{capacity}. And the capacity change over time is displayed in Figure \ref{fig: Capacity vs time} for all cases. Here it is possible to see how the cell can be charged to a higher capacity by limiting the charging current. 

\begin{equation}
    C= \int \mathbf{j} \cdot \mathbf{\hat{n}}  dt
    \label{capacity}
\end{equation}

where $C$ is the capacity of the cell, and  $\hat{n}$ is the unit surface-normal vector at the boundary of the system.

The maximum capacity per unit area of the cell is calculated as in equation \ref{maxCapacity}
\begin{equation}
    C_{max,i}=z_i F c_{0,i} \Delta x
    \label{maxCapacity}
\end{equation}
where component $i$ corresponds to the depositing ion, $c_{0,i}$ is its initial molar concentration and $\Delta x$ is the initial thickness of the electrolyte layer. Under charge operation, this corresponds to $Na^+$, and under discharge mode, this corresponds to the $Zn^{+2}$. Under the current initial conditions, the maximum capacity of the cell under discharge mode is therefore approximately 2128 $Ah/m^2$ and 1829 $Ah/m^2$ for charge mode.

The relation between the maximum capacity and the discharge time may appear to have a discrepancy with the amount of charge that should theoretically have been passed through the cell. This can be explained by the combined contribution of the diffusion current, and the amount of current that remains in the cell after it has reached its mass transfer limit. 

During discharge, the cell appears to behave similarly regardless of whether it is discharged under constant current or constant potential conditions, as it reaches similar discharge limits.

\begin{figure}
    \centering
    \begin{tabular}{l l}
    a) Charge & b) Discharge \\
    \includegraphics[width=7cm]{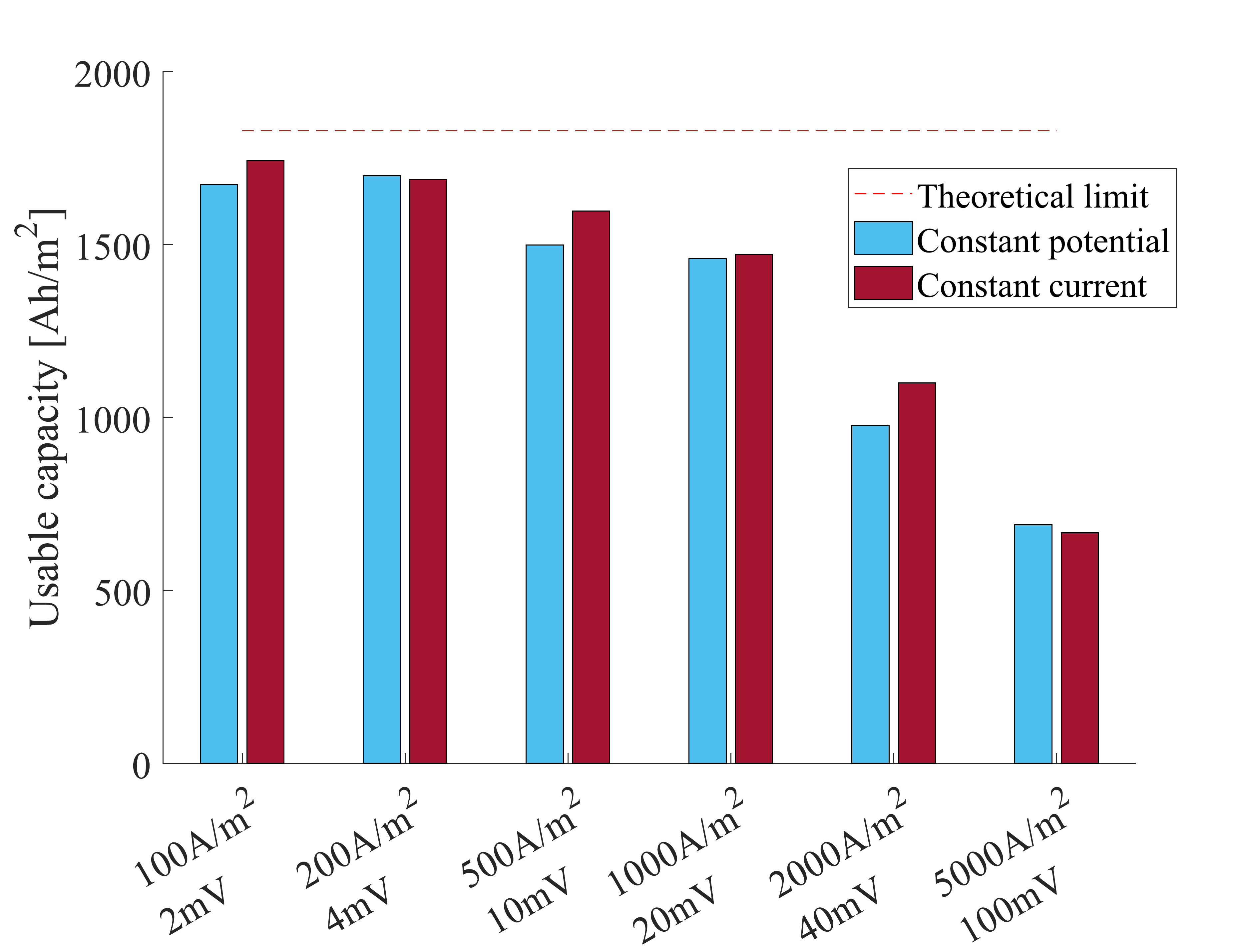}
&    \includegraphics[width=7cm]{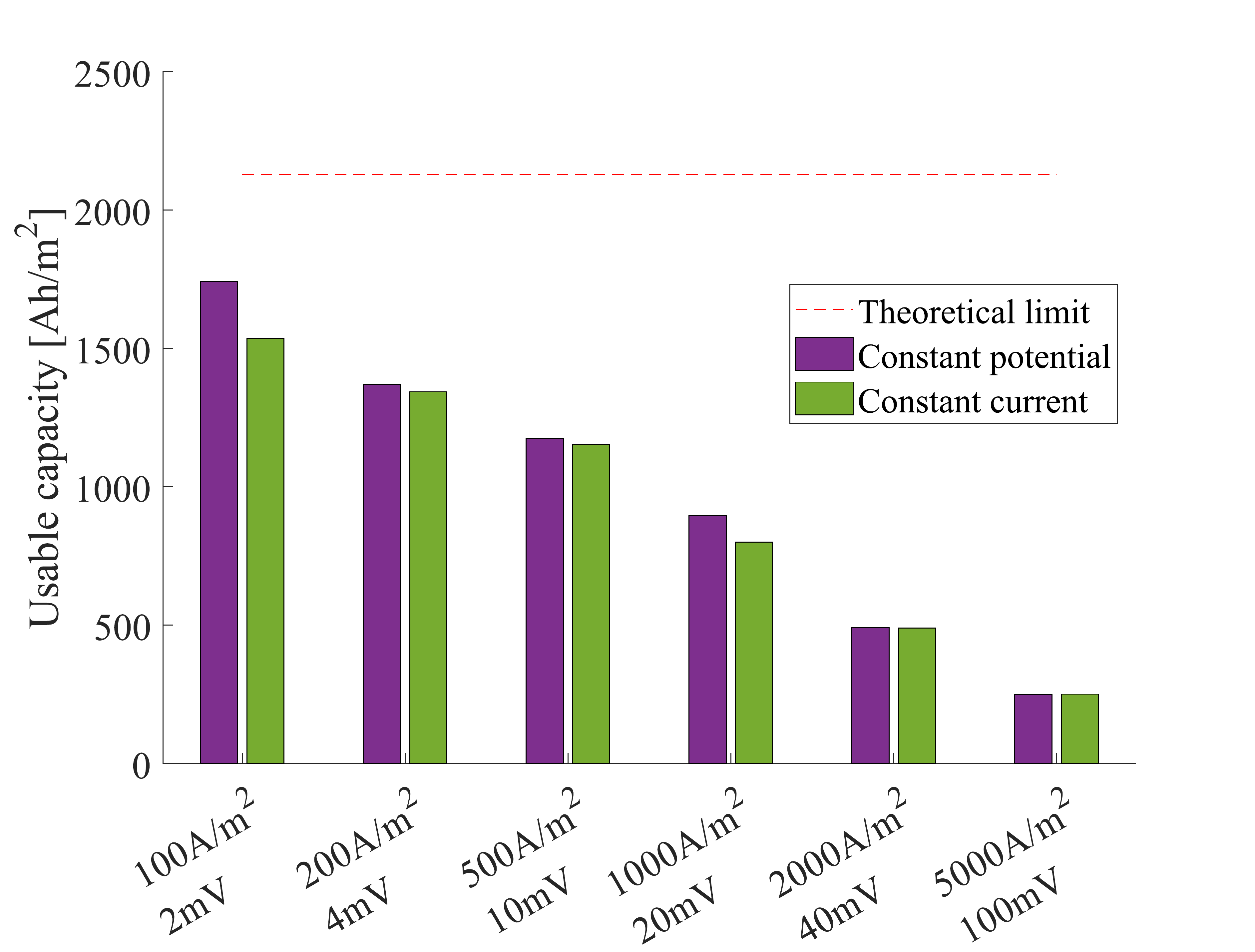}
\end{tabular}
    \caption{Chart of usable capacity for each case under charge (a) and discharge (b), the maximum theoretical capacity that could be charged or discharged is shown in red dashes.}
    \label{fig:usable capacity}
\end{figure}

The maximum usable capacity for each case is shown in Figure \ref{fig:usable capacity}
, where the impact in capacity loss as a result of fast cycling can be clearly observed. In our production, the effect during discharge seems to be somewhat more pronounced.

\begin{figure}
    \centering
        \begin{tabularx}{\textwidth}{c  X X}
       \hspace{1cm} &  Constant current & Constant potential \\
        
        Charge & \includegraphics[width=5cm]{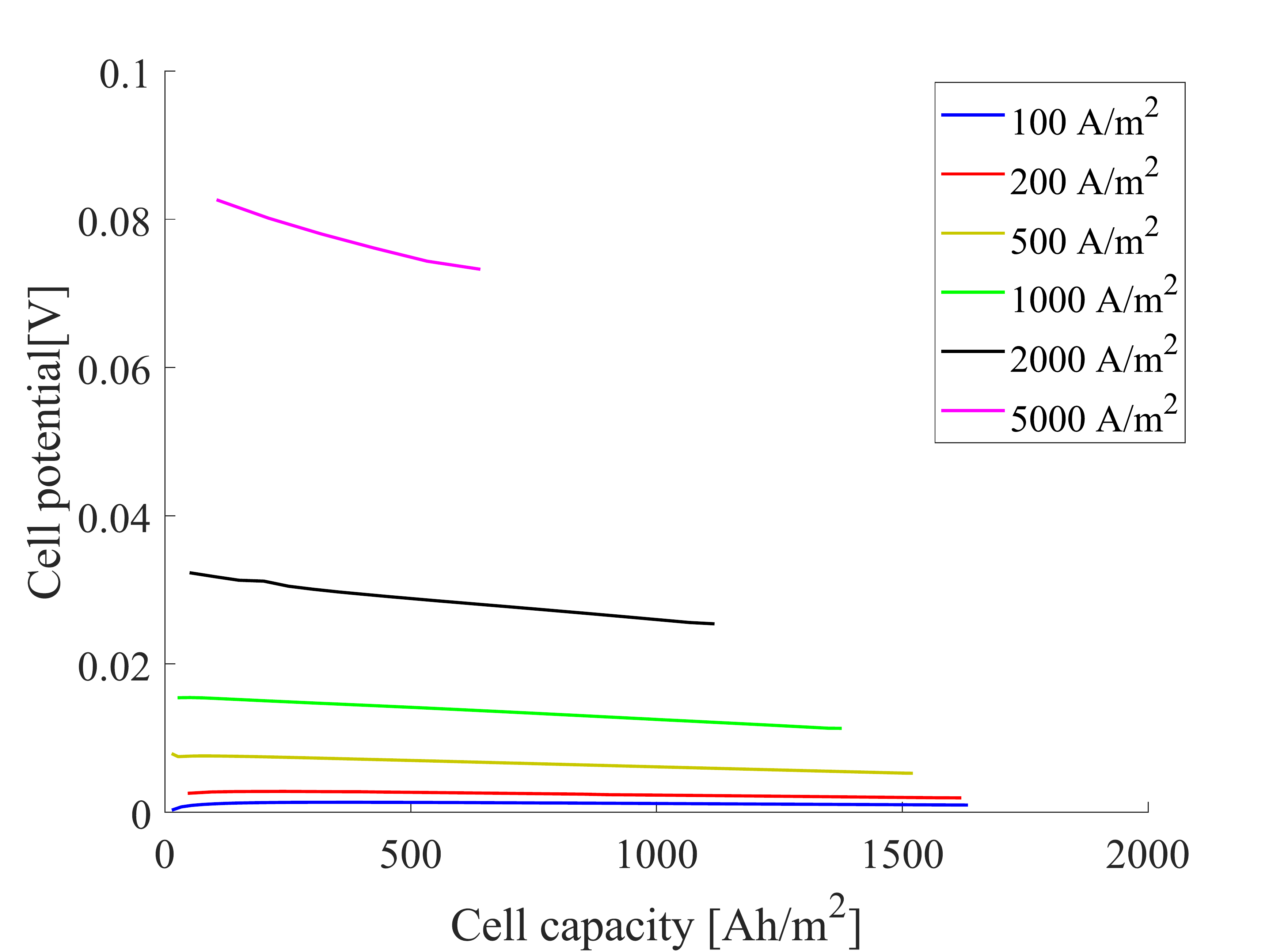} &  \includegraphics[width=5cm]{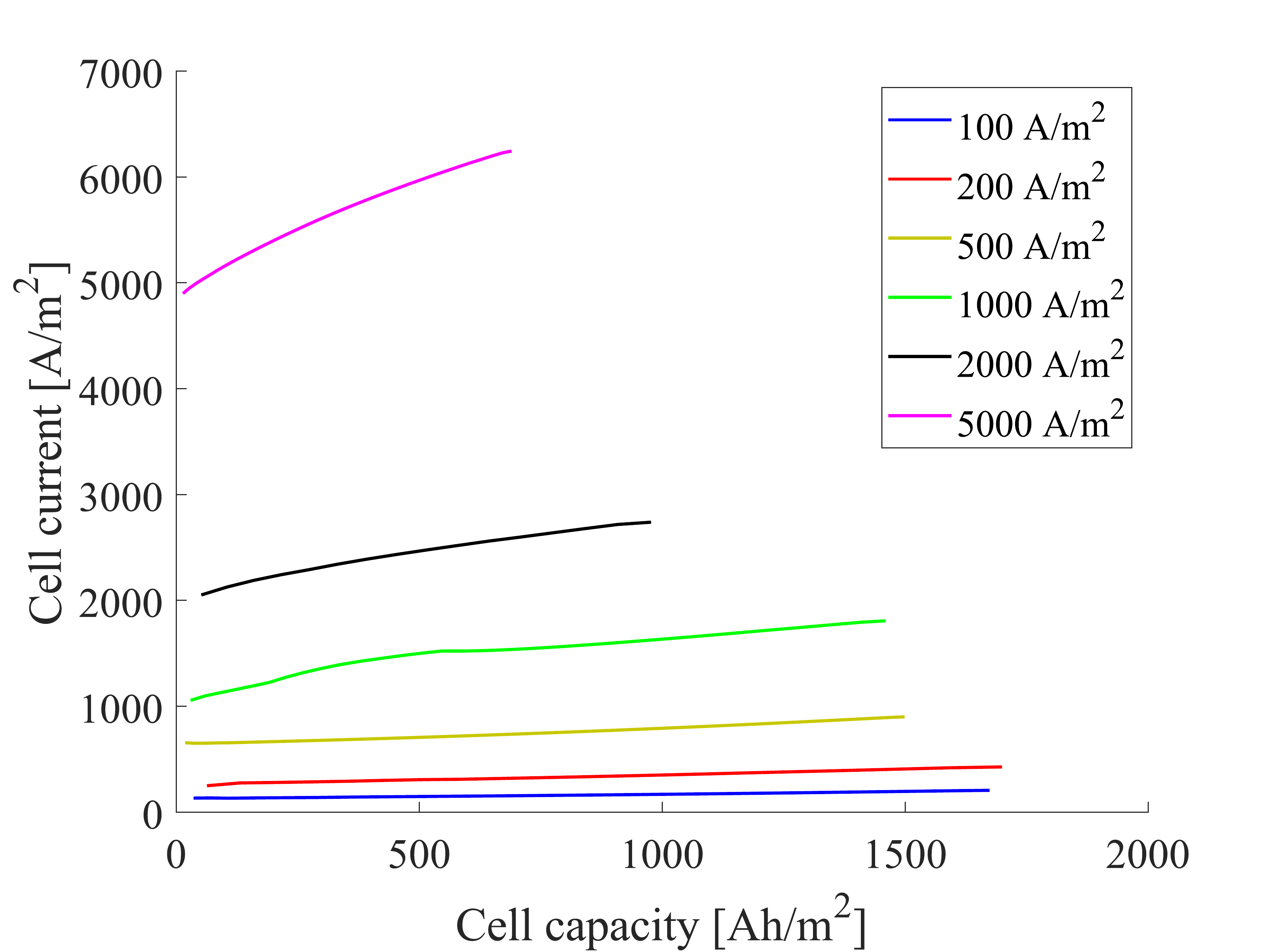} \\
       
        Discharge & \includegraphics[width=5cm]{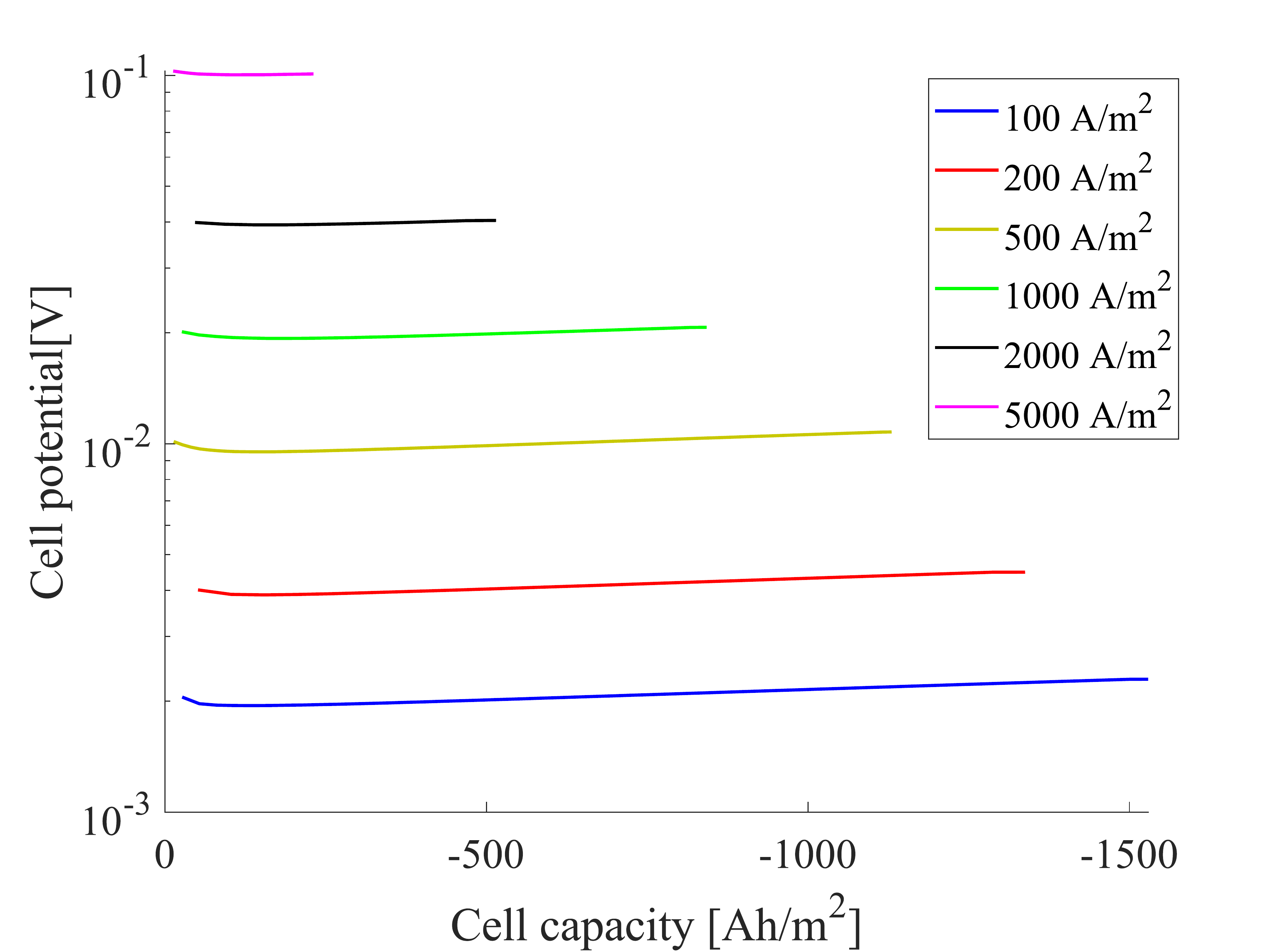} & \includegraphics[width=5cm]{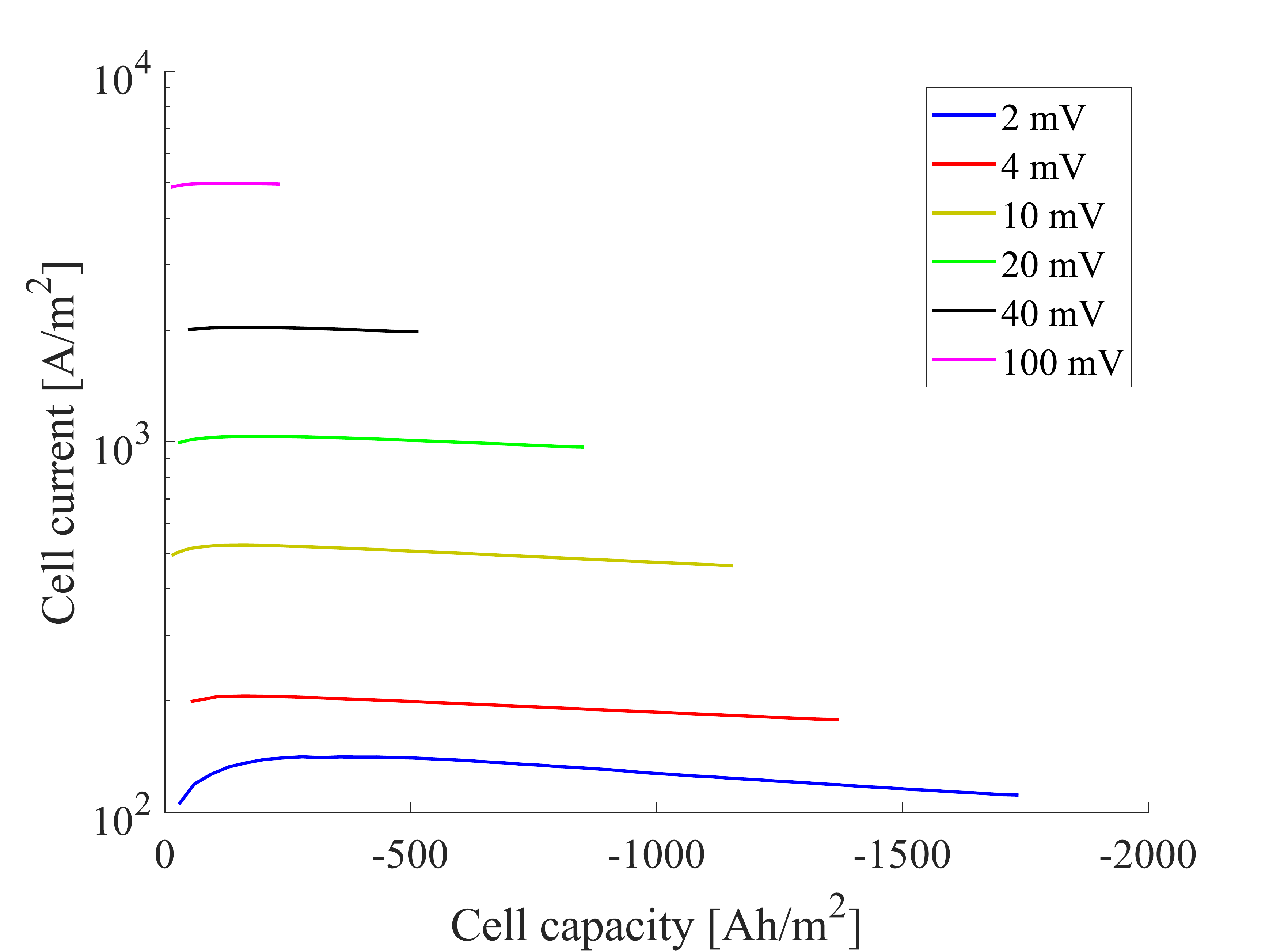}
    \end{tabularx}
    \caption{Cell current and cell voltage for each case against capacity.}
    \label{fig:cell current and potential}
\end{figure}

In Figure \ref{fig:cell current and potential} the corresponding cell potential difference achieved under constant current conditions are shown in the left column, while the right column shows the cell current variation under constant potential. The constant current cases show a smooth decline in the potential  during charge, showing the thinning of the electrolyte layer is the dominant factor. When charging the cell under constant potential, predictably the cell current increases as the total cell resistance is lower.
Under discharge, the potential does not appear to change much under constant current, presumably because of the electrolyte thickness increase offsetting the effect of the diffusion current, since it has been observed that electrolyte thickness has large impact on cell potential \cite{jiang2019effects}. Under constant potential discharge, however, the current seems to vary non-monotonically with a small initial increase and then a decrease over time. This may also be an effect of the initial separation of the components, causing an anisotropic conductivity in the electrolyte layer, and then being offset by the increased resistance associated with the electrolyte expansion.

From these observations, we can speculate that the interplay between the diffusion current and the volumetric changes in the cell may be an important aspect when defining the operation limits and cell size of a sodium-zinc liquid metal battery. Significant differences in usable capacity can be determined by the interaction between these variables and therefore it should be considered when determining the optimal size and operation range of a battery design. 

It must be noted that these observations are made are caused by the compositional variation within the cell, but this would only suggest that these effects may be very important depending on the dimensions of the cell.  The accumulation of Zn ion near the Na electrode, leading to parasitic discharge, highlights the importance of the introduction of a diffusion barrier to mitigate this effect such as the one introduced by Xu et al. \cite{Xu2017a}, however, one must consider that this appears in these results in the absence of convective effects, and thus additional mixing would mitigate this to some extent. 

\section{Conclusions}

Using a simplified multiphase approach, it was possible to model the volumetric changes taking place as a result of the charge and discharge of a Na-Zn liquid metal battery. Significant changes in volume are observed due to the relatively low density of sodium compared to the molten salt electrolyte leading to significant variation in the electrolyte thickness and therefore in the potential distribution across the cell. 

It was shown that concentration gradients in the molten salt composition can lead to the depletion of the reacting ions near the electrode surface and therefore operation limits must consider these concentration gradients. The importance of this electrolyte compositional variation has been highlighted before, but its impact joined with the volumetric changes in the electrolyte leads to a more complex phenomenon.  

It is however noted that these phenomena might be partially mitigated by other transport phenomena that have been neglected under our settings, such as compositional convection and convective effects. 

Future work will address the linkage between chemical composition variation and hydrodynamic effects. The introduction of separator membranes and its impact on species distribution in the electrolyte is also an important factor that should be studied. 

Additional phenomena that cannot be resolved unidimensionally should also be considered, in the future, such as the effect of having non-planar electrode-electrolyte interfaces, and other electro-capillary phenomena, given the high interfacial tension between these materials.  

\section*{Acknowledgement}
This project received funding from the European Union’s Horizon 2020 research and innovation programme under grant agreement No 963599.

%\nocite{*}


%\section{Sample Appendix Section}
 \bibliographystyle{elsarticle-num} 
 \bibliography{1Dpaper_references}

\begin{thebibliography}{10}
\expandafter\ifx\csname url\endcsname\relax
  \def\url#1{\texttt{#1}}\fi
\expandafter\ifx\csname urlprefix\endcsname\relax\def\urlprefix{URL }\fi
\expandafter\ifx\csname href\endcsname\relax
  \def\href#1#2{#2} \def\path#1{#1}\fi

\bibitem{Zhang2021}
S.~Zhang, Y.~Liu, Q.~Fan, C.~Zhang, T.~Zhou, K.~Kalantar-Zadeh, Z.~Guo, Liquid
  metal batteries for future energy storage\href
  {https://doi.org/10.1039/d1ee00531f} {\path{doi:10.1039/d1ee00531f}}.

\bibitem{weier_liquid_2017}
T.~Weier, A.~Bund, W.~El-Mofid, G.~M. Horstmann, C.-C. Lalau, S.~Landgraf,
  M.~Nimtz, M.~Starace, F.~Stefani, N.~Weber, Liquid metal batteries -
  materials selection and fluid dynamics 228  012013, publisher: {IOP}
  Publishing.
\newblock \href {https://doi.org/10.1088/1757-899X/228/1/012013}
  {\path{doi:10.1088/1757-899X/228/1/012013}}.

\bibitem{hadjipaschalis2009overview}
I.~Hadjipaschalis, A.~Poullikkas, V.~Efthimiou, Overview of current and future
  energy storage technologies for electric power applications, Renewable and
  sustainable energy reviews 13~(6-7) (2009) 1513--1522.

\bibitem{Solheim2017}
A.~Solheim, K.~S. Osen, C.~Sommerseth, O.~E. Kongstein, Liquid metal batteries
  as a power buffer in aluminium production plants~(2016)  2--5.

\bibitem{Kim2013}
H.~Kim, D.~A. Boysen, J.~M. Newhouse, B.~L. Spatocco, B.~Chung, P.~J. Burke,
  D.~J. Bradwell, K.~Jiang, A.~A. Tomaszowska, K.~Wang, W.~Wei, L.~A. Ortiz,
  S.~A. Barriga, S.~M. Poizeau, D.~R. Sadoway, Liquid metal batteries: Past,
  present, and future 113~(3)  2075--2099.
\newblock \href {https://doi.org/10.1021/cr300205k}
  {\path{doi:10.1021/cr300205k}}.

\bibitem{li_liquid_2016}
H.~Li, H.~Yin, K.~Wang, S.~Cheng, K.~Jiang, D.~R. Sadoway, Liquid metal
  electrodes for energy storage batteries 6~(14)  1600483, \_eprint:
  https://onlinelibrary.wiley.com/doi/pdf/10.1002/aenm.201600483.
\newblock \href {https://doi.org/10.1002/aenm.201600483}
  {\path{doi:10.1002/aenm.201600483}}.

\bibitem{bradwell2010liquid}
D.~Bradwell, D.~A. Boysen, L.~Ortiz, D.~R. Sadoway, Liquid metal battery: An
  electrometallurgical approach to large-scale energy storage, no.~4, IOP
  Publishing, 2010, p. 187.

\bibitem{ding_room-temperature_2020}
Y.~Ding, X.~Guo, Y.~Qian, L.~Xue, A.~Dolocan, G.~Yu, Room-temperature
  all-liquid-metal batteries based on fusible alloys with regulated interfacial
  chemistry and wetting 32~(30)  2002577, \_eprint:
  https://onlinelibrary.wiley.com/doi/pdf/10.1002/adma.202002577.
\newblock \href {https://doi.org/10.1002/adma.202002577}
  {\path{doi:10.1002/adma.202002577}}.

\bibitem{yin_faradaically_2018}
H.~Yin, B.~Chung, F.~Chen, T.~Ouchi, J.~Zhao, N.~Tanaka, D.~R. Sadoway,
  Faradaically selective membrane for liquid metal displacement batteries 3~(2)
   127--131, number: 2 Publisher: Nature Publishing Group.
\newblock \href {https://doi.org/10.1038/s41560-017-0072-1}
  {\path{doi:10.1038/s41560-017-0072-1}}.

\bibitem{herreman2020solutal}
W.~Herreman, S.~B{\'e}nard, C.~Nore, P.~Personnettaz, L.~Cappanera, J.-L.
  Guermond, Solutal buoyancy and electrovortex flow in liquid metal batteries,
  Physical Review Fluids 5~(7) (2020) 074501.

\bibitem{personnettaz2018thermally}
P.~Personnettaz, P.~Beckstein, S.~Landgraf, T.~K{\"o}llner, M.~Nimtz, N.~Weber,
  T.~Weier, Thermally driven convection in li|| bi liquid metal batteries,
  Journal of Power Sources 401 (2018) 362--374.

\bibitem{Xu2017a}
J.~Xu, A.~M. Martinez, K.~S. Osen, O.~S. Kjos, O.~E. Kongstein, G.~M. Haarberg,
  Electrode behaviors of na-zn liquid metal battery 164~(12)  A2335--A2340.
\newblock \href {https://doi.org/10.1149/2.0591712jes}
  {\path{doi:10.1149/2.0591712jes}}.

\bibitem{Newhouse2014}
J.~M. Newhouse, Modeling the operating voltage of liquid metal battery cells
  by.

\bibitem{Ning2015}
X.~Ning, S.~Phadke, B.~Chung, H.~Yin, P.~Burke, D.~R. Sadoway, Self-healing
  li-bi liquid metal battery for grid-scale energy storage 275  370--376.
\newblock \href {https://doi.org/10.1016/j.jpowsour.2014.10.173}
  {\path{doi:10.1016/j.jpowsour.2014.10.173}}.

\bibitem{kim_calciumbismuth_2013}
H.~Kim, D.~A. Boysen, T.~Ouchi, D.~R. Sadoway, Calcium–bismuth electrodes for
  large-scale energy storage (liquid metal batteries) 241  239--248.
\newblock \href {https://doi.org/10.1016/j.jpowsour.2013.04.052}
  {\path{doi:10.1016/j.jpowsour.2013.04.052}}.

\bibitem{Guo2021}
Z.~Guo, Y.~Zhang, Y.~He, H.~Li, Y.~Wang, K.~Wang, K.~Jiang, Thermal power
  characteristics of a liquid metal battery 7  1221--1230, publisher: Elsevier.
\newblock \href {https://doi.org/10.1016/J.EGYR.2021.09.141}
  {\path{doi:10.1016/J.EGYR.2021.09.141}}.

\bibitem{xie2020high}
H.~Xie, H.~Zhao, J.~Wang, P.~Chu, Z.~Yang, C.~Han, Y.~Zhang, High-performance
  bismuth-gallium positive electrode for liquid metal battery, Journal of Power
  Sources 472 (2020) 228634.

\bibitem{ding2020room}
Y.~Ding, X.~Guo, Y.~Qian, L.~Xue, A.~Dolocan, G.~Yu, Room-temperature
  all-liquid-metal batteries based on fusible alloys with regulated interfacial
  chemistry and wetting, Advanced Materials 32~(30) (2020) 2002577.

\bibitem{entr2014report}
D.~Entr, Report on critical raw materials for the eu, European Commission:
  Brussels, Belgium (2014).

\bibitem{Xu2016}
J.~Xu, O.~S. Kjos, K.~S. Osen, A.~M. Martinez, O.~E. Kongstein, G.~M. Haarberg,
  Na-zn liquid metal battery 332  274--280, publisher: Elsevier B.V.
\newblock \href {https://doi.org/10.1016/j.jpowsour.2016.09.125}
  {\path{doi:10.1016/j.jpowsour.2016.09.125}}.

\bibitem{zhang_anode_2022}
F.~Zhang, J.~Jin, J.~Xu, Z.~Shi, Anode reaction mechanisms of
  na{\textbar}{NaCl}-{CaCl}2{\textbar}zn liquid metal battery 72  81--87.
\newblock \href {https://doi.org/10.1016/j.jechem.2022.04.035}
  {\path{doi:10.1016/j.jechem.2022.04.035}}.

\bibitem{Sun2019}
H.~Sun, G.~Zhu, X.~Xu, M.~Liao, Y.~Y. Li, M.~Angell, M.~Gu, Y.~Zhu, W.~H. Hung,
  J.~Li, Y.~Kuang, Y.~Meng, M.~C. Lin, H.~Peng, H.~Dai, A safe and
  non-flammable sodium metal battery based on an ionic liquid electrolyte
  10~(1), publisher: Nature Publishing Group.
\newblock \href {https://doi.org/10.1038/s41467-019-11102-2}
  {\path{doi:10.1038/s41467-019-11102-2}}.

\bibitem{Gong2020}
Q.~Gong, W.~Ding, A.~Bonk, H.~Li, K.~Wang, A.~Jianu, A.~Weisenburger, A.~Bund,
  T.~Bauer, Molten iodide salt electrolyte for low-temperature low-cost
  sodium-based liquid metal battery 475.
\newblock \href {https://doi.org/10.1016/j.jpowsour.2020.228674}
  {\path{doi:10.1016/j.jpowsour.2020.228674}}.

\bibitem{liu_molten_2022}
H.~Liu, X.~Zhang, S.~He, D.~He, Y.~Shang, H.~Yu, Molten salts for rechargeable
  batteries\href {https://doi.org/10.1016/j.mattod.2022.09.005}
  {\path{doi:10.1016/j.mattod.2022.09.005}}.

\bibitem{zhou_sodium_2022}
H.~Zhou, H.~Li, Q.~Gong, S.~Yan, X.~Zhou, S.~Liang, W.~Ding, Y.~He, K.~Jiang,
  K.~Wang, A sodium liquid metal battery based on the multi-cationic
  electrolyte for grid energy storage 50  572--579.
\newblock \href {https://doi.org/10.1016/j.ensm.2022.05.032}
  {\path{doi:10.1016/j.ensm.2022.05.032}}.

\bibitem{weber_cell_2022}
N.~Weber, C.~Duczek, G.~M. Horstmann, S.~Landgraf, M.~Nimtz, P.~Personnettaz,
  T.~Weier, D.~R. Sadoway, Cell voltage model for li-bi liquid metal batteries
  309  118331.
\newblock \href {https://doi.org/10.1016/j.apenergy.2021.118331}
  {\path{doi:10.1016/j.apenergy.2021.118331}}.

\bibitem{swinkels1971molten}
D.~Swinkels, Molten salt batteries and fuel cells, Advances in Molten Salt
  Chemistry: Volume 1 (1971) 165--223.

\bibitem{herreman_solutal_2020}
W.~Herreman, S.~Bénard, C.~Nore, P.~Personnettaz, L.~Cappanera, J.-L.
  Guermond, Solutal buoyancy and electrovortex flow in liquid metal batteries
  5~(7)  074501, publisher: American Physical Society.
\newblock \href {https://doi.org/10.1103/PhysRevFluids.5.074501}
  {\path{doi:10.1103/PhysRevFluids.5.074501}}.

\bibitem{personnettaz_mass_2019}
P.~Personnettaz, S.~Landgraf, M.~Nimtz, N.~Weber, T.~Weier, Mass transport
  induced asymmetry in charge/discharge behavior of liquid metal batteries 105
  106496.
\newblock \href {https://doi.org/10.1016/j.elecom.2019.106496}
  {\path{doi:10.1016/j.elecom.2019.106496}}.

\bibitem{personnettaz2021effects}
P.~Personnettaz, S.~Landgraf, M.~Nimtz, N.~Weber, T.~Weier, Effects of current
  distribution on mass transport in the positive electrode of a liquid metal
  battery, arXiv preprint arXiv:2104.00144 (2021).

\bibitem{zikanov_metal_2015}
O.~Zikanov, Metal pad instabilities in liquid metal batteries 92~(6)  063021,
  publisher: American Physical Society.
\newblock \href {https://doi.org/10.1103/PhysRevE.92.063021}
  {\path{doi:10.1103/PhysRevE.92.063021}}.

\bibitem{herreman_numerical_2019}
W.~Herreman, C.~Nore, P.~Ziebell~Ramos, L.~Cappanera, J.-L. Guermond, N.~Weber,
  Numerical simulation of electrovortex flows in cylindrical fluid layers and
  liquid metal batteries 4~(11)  113702, publisher: American Physical Society.
\newblock \href {https://doi.org/10.1103/PhysRevFluids.4.113702}
  {\path{doi:10.1103/PhysRevFluids.4.113702}}.

\bibitem{stefani_magnetohydrodynamic_2016}
F.~Stefani, V.~Galindo, C.~Kasprzyk, S.~Landgraf, M.~Seilmayer, M.~Starace,
  N.~Weber, T.~Weier, Magnetohydrodynamic effects in liquid metal batteries 143
   012024, publisher: {IOP} Publishing.
\newblock \href {https://doi.org/10.1088/1757-899X/143/1/012024}
  {\path{doi:10.1088/1757-899X/143/1/012024}}.

\bibitem{Herreman2015}
W.~Herreman, C.~Nore, L.~Cappanera, J.~L. Guermond, Tayler instability in
  liquid metal columns and liquid metal batteries 771  79--114.
\newblock \href {https://doi.org/10.1017/jfm.2015.159}
  {\path{doi:10.1017/jfm.2015.159}}.

\bibitem{Kelley2018}
D.~H. Kelley, T.~Weier, Fluid mechanics of liquid metal batteries 70~(2).
\newblock \href {https://doi.org/10.1115/1.4038699}
  {\path{doi:10.1115/1.4038699}}.

\bibitem{horstmann_coupling_2018}
G.~M. Horstmann, N.~Weber, T.~Weier, Coupling and stability of interfacial
  waves in liquid metal batteries 845  1--35, publisher: Cambridge University
  Press.
\newblock \href {https://doi.org/10.1017/jfm.2018.223}
  {\path{doi:10.1017/jfm.2018.223}}.

\bibitem{benard_anode-metal_2021}
S.~Bénard, N.~Weber, G.~M. Horstmann, S.~Landgraf, T.~Weier, Anode-metal drop
  formation and detachment mechanisms in liquid metal batteries 510  230339.
\newblock \href {https://doi.org/10.1016/j.jpowsour.2021.230339}
  {\path{doi:10.1016/j.jpowsour.2021.230339}}.

\bibitem{Weber2013}
N.~Weber, V.~Galindo, F.~Stefani, T.~Weier, T.~Wondrak, Numerical simulation of
  the tayler instability in liquid metals, New Journal of Physics 15~(4) (2013)
  043034.

\bibitem{weber_numerical_2020}
N.~Weber, M.~Nimtz, P.~Personnettaz, T.~Weier, D.~Sadoway, Numerical simulation
  of mass transfer enhancement in liquid metal batteries by means of
  electro-vortex flow 1  100004.
\newblock \href {https://doi.org/10.1016/j.powera.2020.100004}
  {\path{doi:10.1016/j.powera.2020.100004}}.

\bibitem{duczek2023simulation}
C.~Duczek, N.~Weber, O.~E. Godinez-Brizuela, T.~Weier, Simulation of potential
  and species distribution in a li|| bi liquid metal battery using coupled
  meshes, Electrochimica Acta 437 (2023) 141413.

\bibitem{jiang2019effects}
Y.~Jiang, T.~Cao, P.~Song, D.~Zhang, Y.~Shi, N.~Cai, Effects of magnetically
  induced flow on electrochemical reacting processes in a liquid metal battery,
  Journal of Power Sources 438 (2019) 226926.

\bibitem{newman2021electrochemical}
J.~Newman, N.~P. Balsara, Electrochemical systems, John Wiley \& Sons, 2021.

\bibitem{janz1975molten}
G.~J. Janz, R.~Tomkins, C.~Allen, J.~Downey~Jr, G.~Garner, U.~Krebs, S.~K.
  Singer, Molten salts: volume 4, part 2, chlorides and mixtures—electrical
  conductance, density, viscosity, and surface tension data, Journal of
  Physical and Chemical Reference Data 4~(4) (1975) 871--1178.

\bibitem{sobolev2011database}
V.~Sobolev, Database of thermophysical properties of liquid metal coolants for
  gen-iv (2011).

\bibitem{Weber2019}
N.~Weber, S.~Landgraf, K.~Mushtaq, M.~Nimtz, P.~Personnettaz, T.~Weier,
  J.~Zhao, D.~Sadoway, Modeling discontinuous potential distributions using the
  finite volume method, and application to liquid metal batteries,
  Electrochimica Acta 318 (2019) 857--864.

\bibitem{greenshieldsweller2022}
C.~Greenshields, H.~Weller, Notes on Computational Fluid Dynamics: General
  Principles, CFD Direct Ltd, Reading, UK, 2022.

\end{thebibliography}
 \newpage
\appendix
%\input{Supplementary_information}




% \end{thebibliography}
\end{document}